\documentclass[preprint,showpacs,preprintnumbers,amsmath,amssymb]{revtex4}
\usepackage{graphicx}
\usepackage{floatflt}
\usepackage{dcolumn}
\usepackage{bm}
\usepackage{rotating}

\begin{document}

\preprint{APS/123-QED}
\title{ Non-rotating and rotating neutron stars in the extended field
theoretical model}

\author{Shashi K. Dhiman$^1$, Raj Kumar$^1$ and B. K. Agrawal$^2$  }
\affiliation{
$^1$Department of Physics,
H.P. University, Shimla -  171005,
India.\\
$^2$Saha Institute of Nuclear Physics, Kolkata - 700064, India.}

\begin{abstract}

We study the properties of non-rotating and rotating neutron stars
for a new set of equations of state (EOSs) with different high density
behaviour obtained using the extended field theoretical model.  The high
density behaviour for these EOSs are varied by varying the $\omega-$meson
self-coupling and hyperon-meson couplings in such a way that the quality
of fit to the  bulk nuclear observables, nuclear matter incompressibility
coefficient and hyperon-nucleon potential depths remain practically
unaffected.  We find that the largest value for maximum mass for the
non-rotating neutron star is $2.1M_\odot$.  The radius for the neutron
star with canonical mass is $12.8 - 14.1$ km provided only those  EOSs are
considered for which maximum mass is larger than $1.6M_\odot$ as it is the
lower bound on the maximum mass measured so far.  Our results for the very
recently discovered  fastest rotating  neutron star indicate that this
star is supra massive with mass $1.7 - 2.7M_\odot$ and circumferential
equatorial radius  $12 - 19$ km.

\end{abstract}
 \maketitle

\section{Introduction}
\label{intro_sec}

The knowledge of neutron star properties is necessary to probe the high
density behaviour of the equation of state (EOS) for the baryonic matter
in $\beta-$equilibrium.  The EOS for the densities higher than $\rho_0 =
0.16\text{ fm}^{-3}$ can be well constrained if radii for the neutron
stars over a wide range of their  masses are appropriately known.
Even the accurate information on the maximum neutron star mass
$M_{\text{max}}$ and radius $R_{1.4}$ for the neutron star with canonical
mass ($1.4M_\odot$) would narrow down the choices for the plausible EOSs
to just a few.  Till date, the  neutron stars with masses only around
$1.4M_\odot$ are accurately measured \cite{Thorsett99,Burgay03,Lyne04}.
Recent measurement of mass of the pulsar PSR J0751$+$1807
imposes  lower bounds on the maximum mass of the neutron star to be
$1.6M_\odot$ and $1.9M_\odot$ with $95\%$ and $68\%$ confidence limits,
respectively \cite{Nice05a}.  The increase in the lower bounds of the
neutron star maximum mass could eliminate the family of EOSs in which
exotica appear and substantial softening begins around 2 to 4 $\rho_{0}$
leading to appreciable reduction of  the maximum mass.  The available data
on the  neutron star radius have large uncertainties \cite{Rutledge02,
Rutledge02a,Gendre03,Becker03,Cottam02}. The
main source of the uncertainties in the measurements of the neutron
star radii are the unknown chemical composition of the atmosphere,
inaccuracies in the star's distance and  high magnetic field ($\sim
10^{12}$ G).  Recent discovery of the binary neutron star system
PSR J0737-3039A,B \cite{Burgay03} with masses of the individual star
being $1.338M_\odot$ and $1.249M_\odot$ have raised the hope for the
possibility of measuring the moment of inertia  due to the spin-orbit
coupling effects \cite{Lyne04}. It is expected that  a reasonably accurate
value for neutron star radius can be deduced from the moment of inertia
measurements.  Very recent discovery  of the  fastest rotating neutron
star with rotational frequency of  1122 Hz observed in the X-ray transient
XTE J1739-285 \cite{Kaaret06} has placed an additional constrained on
the EOS at very high density \cite{Lavagetto06}.

Theoretically, the mass-radius relationship and compositions
of the neutron stars are studied using various models which
can be broadly grouped into (i) non-relativistic potential
models \cite{Pandharipande75}, (ii) non-relativistic mean-field
models \cite{Chabanat97,Stone03,Mornas05,Agrawal06},
(iii) field theoretical based relativistic mean-field models
(FTRMF) \cite{Prakash95,Glendenning99,Steiner05}
and (iv) Dirac-Brueckner-Hartree-Fock
model \cite{Muther87,Engvik94,Engvik96,Schulze06}.  Each of these models
can yield EOSs with different high density behaviour which is not yet
well constrained. As a result, neutron star properties vary over a wide
range even for the same model.  In this work we shall mainly focus on the
variations in the properties of the neutron stars obtained within the
FTRMF models.  The FTRMF models predict the values of $M_{\text{max}}
= 1.2 - 3.0M_\odot$ and $R_{1.4} = 10 - 16$ km for the non-rotating
neutron stars  \cite{Muller96,Taurines01,Jha06}.  The lower values of
$M_{\text{max}}$ and $R_{1.4}$ correspond to the neutron stars
composed of nucleons and hyperons in $\beta-$equilibrium, where as,
the higher values of $M_{\text{max}}$ and $R_{1.4}$ correspond to the
neutron stars with no hyperons.  We would like to emphasize that not all
the different parameterizations of the FTRMF model, employed to study
the neutron star properties, are  able to reproduce satisfactorily
the basic properties of finite nuclei and nuclear matter at the
saturation density.  For instance, the value of the nuclear matter
incompressibility coefficient which largely controls the low density
behaviour of a EOS very in between  $200 - 360$ MeV for different FTRMF
models. Though, the value of nuclear matter incompressibility
coefficient is  very well constrained to $230 \pm 10$ MeV by the
experimental data on the isoscalar giant monopole resonances in heavy
nuclei \cite{Reinhard99,Youngblood02}.  The variations in the neutron
star properties resulting from the differences  in the high density
behaviour of the different  EOSs can be appropriately studied only if
the low density behaviour for each of these EOSs are constrained using
the experimental data on the bulk properties of the finite nuclei and
nuclear matter at the saturation density.

The extended FTRMF model \cite{Furnstahl96,Furnstahl97,Serot97} includes mixed
and self-coupling terms for the $\sigma$, $\omega$ and $\rho$ mesons.
The $\omega$-meson self-coupling term enables one to vary the high density
behaviour of the EOS without affecting nuclear matter properties at the
saturation density \cite{Muller96}. The mixed interaction terms involving
$\rho$-mesons allow ones to significantly vary the density dependence
of the symmetry energy coefficient  \cite{Horowitz01,Furnstahl02,Sil05}
which plays crucial role in determining cooling mechanism of a neutron
star \cite{Lattimer91}.  Yet, such a versatile version of the FTRMF
model is not fully explored to study the variations in the properties
of the neutron stars resulting mainly from the uncertainties in the high
density behaviour of EOS.  In the present work we use the extended FTRMF
models to obtain a new set of EOSs with different high density behaviour
for the $\beta-$equilibrated matter composed of nucleons and hyperons.
Each of these different EOSs correspond to different choices  for the
$\omega-$meson self-coupling and hyperon-meson couplings which mainly
affects the high density behaviour of a EOS. The remaining parameters
of the model are calibrated using a set of experimental data on the total
binding energy and charge rms radii for a few closed shell nuclei. In our
calibrational procedure we also use the value of neutron-skin thickness
for the $^{208}$Pb nucleus as one of the data. Since, the neutron-skin
thickness is only poorly known, we obtain different parameter sets
for different neutron-skin thickness ranging from $0.16 - 0.28$ fm.
We further restrict the parameters to yield a reasonable value for the
nuclear matter incompressibility coefficient at the saturation density.
We use our EOSs to study the mass-radius relationship and chemical
compositions for non-rotating neutron stars.  For the case of rotating
neutron stars, we present our results for the Keplerian sequences  and
also investigate the variations of mass and circumferential equatorial
radius for the very recently discovered fastest rotating neutron star.

In Sec. \ref{sec:model} we outline very briefly the Lagrangian
density and corresponding energy density for the extended FTRMF model. In
Sec.  \ref{sec:paramet} we present our various parameterizations for
different combinations of a $\omega$-meson self-coupling,  hyperon-meson
couplings and neutron-skin thickness for the $^{208}$Pb nucleus.
In Sec. \ref{sec:nm_fn} we present our results for the nuclear
matter properties at the saturation density. In this section we also
discuss  about the quality of the fits to the finite nuclei for these
parameterizations. In Sec. \ref{sec:ns} we present our results for
the  properties of non rotating neutron stars.  We also generate some
rotating neutron star sequences for which the results are presented
in Sec. \ref{sec:rot}. Finally our main conclusions are presented in
Sec. \ref{sec:conc}.

\section {Extended  Field Theoretical Model}
\label{sec:model}

The effective Lagrangian density for the FTRMF model generally describes
the interactions of the baryons via the exchange of $\sigma$, $\omega$
and $\rho$ mesons. The $\sigma$ and the $\omega$ mesons are responsible
for nuclear binding while $\rho$ meson is required to obtain the correct
value for the empirical symmetry energy. The cubic and quartic terms for
the self-interaction of the $\sigma$-meson are often considered which
significantly improves the value of the nuclear matter incompressibility.
Nevertheless, the value of the nuclear matter incompressibility
coefficient for these models are usually larger in comparison to
their values extracted from the experimental data on the isoscalar
giant monopole resonances. Moreover,  the symmetry energy coefficient
and its density dependence is also somewhat higher relative to the
corresponding empirical estimates. One  can easily overcome these issues
in the extended FTRMF model which includes self and mixed interaction
terms for $\sigma$, $\omega$ and $\rho$ mesons upto the quartic order.
In particular, mixed interaction terms involving rho-meson  field
enables one to vary the density dependence of the symmetry energy
coefficient and the neutron skin thickness in heavy nuclei  over a
wide range without affecting the other properties of finite nuclei
\cite{Furnstahl02,Sil05}.  The contribution from the self interaction
of $\omega$-mesons plays important role in varying the high density
behaviour of the EOS and also prevents instabilities in the calculation
of the EOS \cite{Sugahara94,Muller96}.  On the other hand expectation
value of the $\rho$-meson field is order of magnitude smaller than that
for the $\omega$-meson  field \cite{Serot97}. Thus, inclusion of the
$\rho$-meson self interaction can affect the properties of the finite
nuclei and neutron stars  only very marginally \cite{Muller96}.

The Lagrangian density
for the extended FTRMF model can be written as,
\begin{equation}
\label{eq:lden}
{\cal L}= {\cal L_{BM}}+{\cal L_{\sigma}} + {\cal L_{\omega}} + {\cal L_{\mathbf{\rho}}} + {\cal L_{\sigma\omega\mathbf{\rho}}} + {\cal L}_{em} + {\cal L}_{e\mu} +  L_{YY}.
\end{equation}
Where the baryonic and mesonic Lagrangian ${\cal L_{BM}}$ can be written, 
\begin{equation}
\label{eq:lbm}
{\cal L_{BM}} = \sum_{B} \overline{\Psi}_{B}[i\gamma^{\mu}\partial_{\mu}-(M_{B}-g_{\sigma B} \sigma)-(g_{\omega B}\gamma^{\mu} \omega_{\mu}+\frac{1}{2}g_{\mathbf{\rho}B}\gamma^{\mu}\tau_{B}.\mathbf{\rho}_{\mu})]\Psi_{B}. 
\end{equation}
Here, the sum is taken over the complete baryon octet which
consists of nucleons, $\Lambda, \Sigma$ and $\Xi$ hyperons.  For the
calculation of finite nuclei properties only neutron and proton has been
considered. $\tau_B$ are the isospin matrices. The Lagrangian describing
self interactions for $\sigma$, $\omega$,   and $\rho$ mesons can be
written as,
\begin{equation}
\label{eq:lsig}
{\cal L_{\sigma}} = \frac{1}{2}(\partial_{\mu}\sigma\partial^{\mu}\sigma-m_{\sigma}^2\sigma^2) -\frac{\overline{\kappa}}{3!}
g_{\sigma N}^3\sigma^3-\frac{\overline{\lambda}}{4!}g_{\sigma N}^4\sigma^4,
\end{equation}
\begin{equation}
\label{eq:lome}
{\cal L_{\omega}} = -\frac{1}{4}\omega_{\mu\nu}\omega^{\mu\nu}+\frac{1}{2}m_{\omega}^2\omega_{\mu}\omega^{\mu}+\frac{1}{4!}\zeta g_{\omega N}^{4}(\omega_{\mu}\omega^{\mu})^{2},
\end{equation}
\begin{equation}
\label{eq:lrho}
{\cal L_{\mathbf{\rho}}} = -\frac{1}{4}\mathbf{\rho}_{\mu\nu}\mathbf{\rho}^{\mu\nu}+\frac{1}{2}m_{\rho}^2\mathbf{\rho}_{\mu}\mathbf{\rho}^{\mu}+\frac{1}{4!}\xi g_{\rho N}^{4}(\mathbf{\rho}_{\mu}\mathbf{\rho}^{\mu})^{2}.
\end{equation}
The $\omega^{\mu\nu}$, $\mathbf{\rho}^{\mu\nu}$ are field tensors corresponding to the 
$\omega$ and $\rho$ mesons, and can be defined as 
$\omega^{\mu\nu}=\partial^{\mu}\omega^{\nu}-\partial^{\nu}\omega^{\mu}$ and 
$\mathbf{\rho}^{\mu\nu}=\partial^{\mu}\mathbf{\rho}^{\nu}-\partial^{\nu}\mathbf{\rho}^{\mu}$. 
The mixed interactions of $\sigma, \omega$, and $\mathbf{\rho}$ mesons ${\cal L_{\sigma\omega\rho}}$ can be written as, 
\begin{equation}
\label{eq:lsor}
\begin{split}
{\cal L_{\sigma\omega\rho}} & = g_{\sigma N}g_{\omega N}^2\sigma\omega_{\mu}\omega^{\mu} \left(\overline{\alpha_1}+\frac{1}{2}\overline{\alpha_1}^{\prime}\sigma\right)
+g_{\sigma N}g_{\rho
N}^{2}\sigma\rho_{\mu}\rho^{\mu}\left(\overline{\alpha_2}+\frac{1}{2}\overline{\alpha_2}^{\prime}
\sigma\right)\\
& +\frac{1}{2}\overline{\alpha_3}^{\prime}g_{\omega N}^{2}g_{\rho N}^2\omega_{\mu}\omega^{\mu}\rho_{\mu}\rho^{\mu}
\end{split}
\end{equation}
The ${\cal L}_{em}$ is Lagrangian for electromagnetic interactions and can be expressed as,
\begin{equation}
\label{eq:lem}
{\cal L}_{em}= -\frac{1}{4}F_{\mu\nu}F^{\mu\nu}- \sum_{B}e\overline{\Psi} _{B}\gamma_{\mu}\frac{1+\tau_{3B}}{2}A_{\mu}\Psi_{B},
\end{equation}
where,  $F^{\mu\nu}=\partial^{\mu}A^{\nu}-\partial^{\nu}A^{\mu}$.
The hyperon-hyperon interaction has been included by introducing two
additional mesonic fields ($\sigma^*$ and $\phi$) and the corresponding
Lagrangian ${\cal L}_{YY}$ ($Y = \Lambda, \Sigma$, and $\Xi$) can be
written as,
\begin{equation}
\label{eq:lyy}
\begin{split}
{\cal L}_{YY} & =
\sum_{Y}\overline{\Psi}_{Y}\left(g_{\sigma^{*}Y}\sigma^{*}-g_{\phi
Y}\gamma^{\mu}\phi_{\mu}\right)\Psi_{Y}\\
&+\frac{1}{2}\left(\partial_{\nu}\sigma^*\partial^{\nu}\sigma^*-
m_{\sigma^*}^2\sigma^{*2}\right)
-\frac{1}{4}S_{\mu\nu}S^{\mu\nu}+\frac{1}{2}m_{\phi}^2\phi_{\mu}\phi^{\mu}.
\end{split}
\end{equation}

The charge neutral neutron star matter also includes leptons such as
$e^-$ and $\mu^-$ in addition to neutrons, proton, and hyperons at the
high densities.  The leptonic contributions to the total Lagrangian
density can be written as,
\begin{equation}
\label{eq:lemu}
{\cal L}_{e\mu}=\sum_{\ell=e,\mu}{\overline{\Psi}_{\ell}}\left(i\gamma^{\mu}\partial_{\mu} -
M_{\ell}\right) \Psi_{\ell}. 
\end{equation}

The equation of motion for baryons, mesons and photons can be derived
from the Lagrangian density defined in Eq.(\ref{eq:lden}). The equation
of motion for baryons can be given as,
\begin{equation}
\label{eq:dirac}
\begin{split}
&\bigg[\gamma^\mu\left(i\partial_\mu - g_{\omega
B}\omega_\mu-\frac{1}{2}g_{\rho B}\tau_{B}.\rho_\mu - e
\frac{1+\tau_{3B}}{2}A_\mu - g_{\phi B}\phi^\mu\right) - \\
&  (M_B + g_{\sigma B}\sigma + g_{\sigma^* B}\sigma^*)\bigg]\Psi_B  
=\epsilon_B \Psi_B. 
\end{split}
\end{equation}

The Euler-Lagrange equations for the ground state expectation values of the mesons fields are: 
\begin{equation}
\label{eq:sigma}
\begin{split}
\left(-\Delta + m_{\sigma}^{2}\right)\sigma & = \sum_{B} g_{\sigma B}\rho_{sB} -\frac{ \overline{\kappa}}{2} g_{\sigma N}^{3}\sigma^{2}- \frac{\overline{\lambda}}{6} g_{\sigma N}^{4}\sigma ^{3} + 
\overline{\alpha_{1}} g_{\sigma N} g_{\omega N}^{2}\omega ^{2}\\
& +\overline{\alpha_{1}}^\prime g_{\sigma N}^{2}
 g_{\omega N}^{2}\sigma\omega ^{2}
 + \overline{\alpha_{2}} g_{\sigma N} g_{\rho B}^{2}\rho ^{2}
 + \overline{\alpha_{2}}^\prime g_{\sigma N}^{2} 
g_{\rho N}^{2}\sigma\rho ^{2}, 
\end{split}                                                                    
\end{equation}
\begin{equation}
\label{eq:omega}
\begin{split}
\left(-\Delta + m_{\omega}^{2}\right)\omega & = \sum_{B} g_{\omega B}\rho_{B} 
- \frac{\zeta}{6} g_{\omega N}^{4}\omega ^{3} 
-2 \overline{\alpha_{1}}g_{\sigma N} g_{\omega N}^{2}\sigma\omega 
-\overline{\alpha_{1}}^\prime g_{\sigma N}^{2}
 g_{\omega N}^{2}\sigma^{2}\omega \\
& - \overline{\alpha_{3}}^\prime g_{\omega N}^{2} 
g_{\rho N}^{2}\omega\rho ^{2}, 
\end{split}                                                                    
\end{equation}
\begin{equation}
\label{eq:rho}
\begin{split}
\left(-\Delta + m_{\rho}^{2}\right)\rho & = \sum_{B} g_{\rho B}\tau_{3B}\rho_{B}- \frac{\xi}{6} g_{\rho N}^{4}\rho ^{3} 
-2 \overline{\alpha_{2}}
 g_{\sigma N} g_{\rho N}^{2}\sigma\rho 
-\overline{\alpha_{2}}^\prime g_{\sigma N}^{2}
 g_{\rho N}^{2}\sigma^{2}\rho \\
& - \overline{\alpha_{3}}^\prime g_{\omega N}^{2} 
g_{\rho N}^{2}\omega^{2}\rho,  
\end{split}                                                                    
\end{equation}
\begin{equation}
\label{eq:sigma*}
\left(-\Delta + m_{\sigma^*}^{2}\right)\sigma^{*} = \sum_{B} g_{\sigma^* B}
\rho_{sB}, 
\end{equation}
\begin{equation}
\label{eq:phi}
\left(-\Delta + m_{\phi}^{2}\right)\phi = \sum_{B} g_{\phi B}
\rho_{B}, 
\end{equation}

\begin{equation}
\label{eq:photon}
-\Delta A_{0} = e\rho_{p}.
\end{equation}
where the baryon density $\rho_B$,  scalar density $\rho_{sB}$ and charge
density $\rho_{p}$ are, respectively,
\begin{equation}
\rho_{B}= \left< \overline{\Psi}_B \gamma^0 \Psi_B\right> = \frac{\gamma k_{B}^{3}}{6\pi^{2}},
\end{equation}

\begin{equation}
\rho_{sB} = \left< \overline{\Psi}_B\Psi_B \right> 
          = \frac{\gamma}{(2\pi)^3}\int_{0}^{k_{B}}d^{3}k \frac{M_{B}^*}
            {\sqrt{k^2 + M_{B}^{*2}}},
\end{equation}
\begin{equation}
\rho_{p} = \left< \overline{\Psi}_B\gamma^{0}\frac{1+\tau_{3B}}{2}\Psi_B
\right>.
\end{equation}
Where, $\gamma$ is the spin degeneracy. The $M_{B}^{*} = M_{B}
- g_{\sigma B}\sigma - g_{\sigma^{*}B}\sigma ^{*}$ is the effective mass
of the baryon species B,  $k_{B}$ is its Fermi momentum and $\tau_{3B}$
denotes the isospin projections of baryon B.

The energy density of the uniform matter  in the extended FTRMF models is given by
\begin{equation}
\label{eq:eden}
\begin{split}
{\cal E} & = \sum_{j=B,\ell}\frac{1}{\pi^{2}}\int_{0}^{k_j}k^2\sqrt{k^2+M_{j}^{*2}} dk
+\sum_{B}g_{\omega B}\omega\rho_{B}+\sum_{B}g_{\rho B}\tau_{3B}\rho
+ \frac{1}{2}m_{\sigma}^2\sigma^2\\
&+\frac{\overline{\kappa}}{6}g_{\sigma N}^3\sigma^3
+\frac{\overline{\lambda}}{24}g_{\sigma N}^4\sigma^4
-\frac{\zeta}{24}g_{\omega N}^4\omega^4-\frac{\xi}{24}g_{\rho N}^4\rho^4
 - \frac{1}{2} m_{\omega}^2 \omega ^2
-\frac{1}{2} m_{\rho}^2 \rho ^2\\
&-\overline{\alpha_1} g_{\sigma N}
 g_{\omega N}^{2}\sigma \omega^2-\frac{1}{2} 
\overline{\alpha_1}^\prime g_{\sigma N}^2 g_{\omega N}^2\sigma^2 \omega^2
-\overline{\alpha_2}g_{\sigma N}g_{\rho N}^2 \sigma\rho^2
 -\frac{1}{2} \overline{\alpha_2}^\prime g_{\sigma N}^2 g_{\rho N}^2\sigma^2 
\rho^2\\
& - \frac{1}{2} \overline{\alpha_3}^\prime g_{\omega N}^2 g_{\rho N}^2
\omega^2\rho^2 +\frac{1}{2} m_{\sigma^*}^{2} {\sigma ^*} ^{2}+\sum_{B}g_{\phi B}\phi\rho_{B} -\frac{1}{2} m_{\phi}^{2} {\phi} ^{2}.
\end{split}
\end{equation}
The pressure of the uniform matter  is given by
\begin{equation}
\label{eq:pden}
\begin{split}
P & = \sum_{j=B,\ell}\frac{1}{3\pi^{2}}\int_{0}^{k_j}
\frac{k^{4}dk}{\sqrt{k^2+M_{j}^{*2}}} 
- \frac{1}{2}m_{\sigma}^2\sigma^2-\frac{\overline{\kappa}}{6}g_{\sigma N}^3\sigma^3 -\frac{\overline{\lambda}}{24}g_{\sigma N}^4\sigma^4\\
& +\frac{\zeta}{24}g_{\omega N}^4\omega^4+\frac{\xi}{24}g_{\rho N}^4\rho^4
  + \frac{1}{2} m_{\omega}^2 \omega ^2
+\frac{1}{2} m_{\rho}^2 \rho ^2+\overline{\alpha_1} g_{\sigma N}
g_{\omega N}^{2}\sigma \omega^2\\
&+\frac{1}{2} \overline{\alpha_1}^\prime g_{\sigma N}^2 g_{\omega N}^2\sigma^2 \omega^2+\overline{\alpha_2}g_{\sigma N}g_{\rho N}^2 \sigma\rho^2
 +\frac{1}{2} \overline{\alpha_2}^\prime g_{\sigma N}^2 g_{\rho N}^2\sigma^2 
\rho^2\\ 
& + \frac{1}{2} \overline{\alpha_3}^\prime g_{\omega N}^2 g_{\rho N}^2
\omega^2\rho^2
 -\frac{1}{2} m_{\sigma^*}^{2} {\sigma ^*} ^{2} 
 +\frac{1}{2} m_{\phi}^{2} {\phi} ^{2}.%
\end{split}
\end{equation}

\section{Parameterizations of the extended FTRMF model}
\label{sec:paramet}

In this section we consider various parameterizations of the extended FTRMF
model. The different parameter sets are obtained using different values
for the $\omega$-meson self-coupling $\zeta$, hyperon-meson couplings
$g_{iY}$ ($i = \sigma, \omega, \rho, \sigma^*$ and $\phi$ mesons)
and neutron-skin thickness $\Delta r$ for the $^{208}$Pb nucleus. The
parameter $\zeta$ mainly affects the high density behaviour of the EOS
and can not be well constrained by the properties of the finite
nuclei. The different sets of $g_{iY}$ can be obtained to yield different
EOSs for the dense matter without affecting the resulting potential
depth for hyperons in the nuclear matter at the saturation density.
The value of $\Delta r$ for a single  heavy nucleus like $^{208}$Pb
which can constrain the linear density dependence of the symmetry energy
is only poorly known.  The different choices for the $\zeta$, $g_{iY}$
and $\Delta r$ are so made that they span entire range of values as often
used in the literature. We must point out that the contributions
from the $\rho$-meson self-coupling are ignored, because, their effects
are found to be only marginal even for the pure neutron matter at very
high densities \cite{Muller96}.

Towards our parameterizational procedure we first set hyperon-meson
couplings $g_{iY} = 0$ in Eqs. (\ref{eq:lbm},\ref{eq:lyy}).  Then
the remaining coupling parameters  appearing in Eqs. (\ref{eq:lbm}
- \ref{eq:lsor}) are determined by fitting the FTRMF results to
the experimental data for the total binding energies and charge rms
radii for many closed shell normal and exotic nuclei.  We consider
total binding energies for $^{16,24}$O, $^{40,48}$Ca, $^{56,78}$Ni,
$^{88}$Sr, $^{90}$Zr, $^{100,116,132}$Sn and $^{208}$Pb nuclei, charge
rms radii for $^{16}$O, $^{40,48}$Ca, $^{56}$Ni, $^{88}$Sr, $^{90}$Zr,
$^{116}$Sn and $^{208}$Pb nuclei. In addition, we also fit the value of
neutron-skin thickness for $^{208}$Pb nucleus. Recently extracted value of
neutron-skin thickness for $^{208}$Pb nucleus from the isospin diffusion
data lie within $0.16 - 0.24\text{ fm}$ indicating large uncertainties
\cite{Chen05}.  We generate twenty one different parameter sets using
different combinations of $\zeta$ and $\Delta r$. The value of $\zeta$
is taken to be $0.0, 0.03$ and $0.06$ and for the $\Delta r$ we use $0.16,
0.18,...., 0.28$ fm.  The best fit parameters are obtained by minimizing
the $\chi^2$ function given as,
 \begin{equation} \chi^2 =  \frac{1}{N_d - N_p}\sum_{i=1}^{N_d}
\left (\frac{ {\cal O}_i^{exp} - {\cal O}_i^{th}}{\delta_i}\right )^2 \label {eq:chi2}
\end{equation} 
where, $N_d$ is the number of  experimental data points and $N_p$
the number of parameters to be fitted. The $\delta_i$ stands for
theoretical error and ${\cal O}_i^{exp}$ and ${\cal O}_i^{th}$ are the
experimental and the corresponding theoretical values, respectively, for
a given observable.  Since, the ${\cal O}_i^{th}$ in Eq.(\ref{eq:chi2})
is calculated using the FTRMF model, the value of $\chi^2$ depends on the
values of the  parameters appearing in Eq. (\ref{eq:lbm} - \ref{eq:lsor}).
The theoretical error  $\delta_i $ in Eq.(\ref{eq:chi2}) are taken to be
1.0 MeV for the total binding energies, 0.02 fm for the  charge rms radii
and 0.005 fm for the neutron-skin thickness.  The best fit parameters
for a given set of values of ${\cal O}_i^{exp}$ and $\delta_i$ are
searched using the simulated annealing method \cite{Agrawal05,Kumar06}.
In our earlier work \cite{Kumar06} we have obtained the parameter sets
for $\zeta = 0.0, 0.03$ and $0.06$ with $\Delta r = 0.18$ fm.  Here too
we follow the same strategy to obtain the parameter set for a given
combination of  $\Delta r$ and $\zeta $.  In Tables \ref{tab:para1},
\ref{tab:para2} and \ref{tab:para3} we list the values of parameters
for all the sets presently generated.

We now determine the values of the hyperon-meson coupling parameters
$g_{iY}$.  These couplings can be expressed in terms of
the nucleon-meson couplings using SU(6) model as,
 \begin{eqnarray}
\label{eq:gy_su6} \frac{1}{3}g_{\sigma N} = \frac{1}{2}g_{\sigma\Lambda}
= \frac{1}{2}g_{\sigma\Sigma}= g_{\sigma\Xi}, \nonumber\\
\frac{1}{3}g_{\omega N} = \frac{1}{2}g_{\omega\Lambda} =
\frac{1}{2}g_{\omega\Sigma}= g_{\omega\Xi}, \nonumber\\
g_{\rho N} = g_{\rho\Sigma} = 2 g_{\rho\Xi}, \qquad
g_{\rho\Lambda} = 0, \nonumber\\ 2g_{\sigma^*\Lambda}=
2g_{\sigma^*\Sigma}=g_{\sigma^*\Xi}=\frac{2\sqrt{2}}{3}g_{\omega N},
 \qquad g_{\sigma^* N}=0, \nonumber\\
2g_{\phi\Lambda}= 2g_{\phi\Sigma}=g_{\phi\Xi}=\frac{2\sqrt{2}}{3}g_{\omega
N},
 \qquad g_{\phi N}=0.
\end{eqnarray} The neutron star properties are quite sensitive to the
values of $g_{\sigma Y}$ and $g_{\omega Y}$. Where as neutron star
properties do not get significantly affected even  if the values of
$g_{\sigma^* Y}$ is  varied over a reasonable range for a fixed value of
$g_{\Phi Y}$ \cite{Bednarek05}. For $g_{\rho Y}$, $g_{\sigma^* Y}$ and
$g_{\Phi Y}$ we use the values as given by Eq. (\ref{eq:gy_su6}). The
values of $g_{\sigma Y}$ and $g_{\omega Y}$ are determined using the
expressions for the hyperon-nucleon potential. The potential depth for a
given hyperon species in the  nuclear matter at the saturation density
($\rho_{sat}$) is given as, \begin{equation} U_{Y}^{(N)}(\rho_{sat})
= -g_{\sigma Y}\sigma(\rho_{sat})+g_{\omega Y}\omega(\rho_{sat}).
\label{eq:uyn} \end{equation} The values of $U_Y^{(N)}$ chosen are as
follows \cite{Schaffner00}, \begin{equation} U_{\Lambda}^{(N)}= -28\text{
MeV}, \qquad U_{\Sigma}^{(N)} = +30\text{ MeV} \qquad \text{and} \qquad
U_{\Xi}^{(N)} = -18\text{ MeV}.  \end{equation} Normally, $g_{\sigma
Y}$ is determined for a given value of $U_{Y}^{(N)}(\rho_{sat})$ with
$g_{\omega Y}$ taken from SU(6) model.  For the sake of convenience
we define, \begin{eqnarray} X_{m Y}=\left\{ \begin{array}{cc} \left(
\frac{g_{m Y}}{g_{m N}}\right)& \text{ for $\Lambda$ and $\Sigma$
hyperons}\\ \\ 2\left( \frac{g_{m Y}}{g_{m N}}\right)& \text{ for
$\Xi$  hyperons}, \end{array} \right .  \label{eq:xw} \end{eqnarray}
where, $m$ stands for $\sigma$ and $\omega$ mesons.   In the present
work we vary $X_{\omega Y}$ from $0.5-0.8$ \cite{Glendenning91}. In
Fig. \ref{fig:xsig_xome} we  display the variations of $X_{\sigma Y}$ as a
function of $X_{\omega Y}$  obtained using the parameter set corresponding
to $\zeta = 0.03$ and $\Delta r = 0.22$ fm. The values of $X_{\sigma Y}$
are for all other combinations of $\zeta$ and $\Delta r$ are very much
the same as depicted in Fig. \ref{fig:xsig_xome}. This is due to the fact
that the properties of symmetric nuclear matter, like, binding energy
per nucleon $B/A$, nuclear matter incompressibility coefficient $K$,
effective nucleon mass $M^*_N$ at the saturation density  are very much
similar for  all the parameterizations considered in the present work.

\section{Nuclear matter and Finite nuclei}
\label{sec:nm_fn}
The various properties associated with the nuclear matter are obtained
using parameter sets of Tables \ref{tab:para1}, \ref{tab:para2}
and \ref{tab:para3}.  The values of 
$B/A$, 
$K$, $M^*_N$ and
$\rho_{sat}$ for all these parameter sets lie in a narrow range. We
find that $B/A = 16.11\pm0.04$ MeV, $K=230.24\pm 9.80$ MeV,
$M^*_N /M_N=0.605\pm0.004$ and $\rho_{sat} = 0.148 \pm0.003$ fm$^{-3}$.
The values of the symmetry energy coefficient $J$ and its linear density
dependence,

\begin{equation}
\label{eq:L}
L=\left . 3\rho\frac{dJ}{d\rho}\right |_{\rho_{sat}}
\end{equation} 
are strongly correlated with the $\Delta r$ for the $^{208}$Pb nucleus
used in the fit. In Fig. \ref{fig:jl_skin} we display the variations of
$J$ and $L$ calculated at  saturation density as a function of $\Delta r$.
The values of  $L$ lie in the range of $80\pm 20$ MeV for $\Delta r $ varying
in between $0.16$ to $0.28$ fm which is in reasonable agreement with the
recent predictions based on the isospin diffusion data \cite{Chen05}.

The relative errors in the total  binding
energy and  charge rms  radius for the nuclei included in the fits
are more or less the same as we have obtained
in our earlier work \cite{Kumar06}.  So, we do not wish to present
here the detailed results. It might be sufficient for the present
purpose to display the results for the rms errors for the total binding
energies and  charge rms  radii obtained for our newly generated parameter
sets. In Fig. \ref{fig:error_br} we plot the rms errors for the total
binding energies and charge radii as a function of $\Delta r$. It is
quite clear from this figure that rms error show hardly any variations
implying that all the parameter sets generated in the present work
fit the finite nuclear properties equally well.  In fact, if we do not
consider the parameterizations with $\zeta = 0.0$ and $\Delta r = 0.26$
or 0.28 fm, the rms errors on the total binding energy are 1.5 - 1.8
MeV which is comparable with one obtained using NL3 parameterizations as
most commonly used \cite{Lalazissis97}.  The rms error of charge radii
for the nuclei considered in the fit lie within the  $0.025 - 0.040$ fm.

\section{Non-rotating Neutron stars}
\label{sec:ns}
In this section we present our results for the properties of
the non-rotating neutron stars for a set of EOSs obtained  using
different parameterizations for the extended FTRMF model. Each of
these parameterizations  corresponds to different combinations of
neutron-skin thickness $\Delta r$ in $^{208}$Pb nucleus, the $\omega$
-meson self-coupling $\zeta$ and hyperon-meson couplings $X_{\omega
Y}$ as described in Sec. \ref{sec:paramet}. The values of $\Delta r$,
$\zeta$ and $X_{\omega Y}$ are so varied that they span the entire range
of values  as often encountered in the literature. The variations in
$\zeta$ and $X_{\omega Y}$ affect the high density behaviour of the EOS,
whereas, the density dependence of the symmetry energy coefficient is
strongly correlated with $\Delta r$.  It is therefore  natural to expect that
the variations in  $\Delta r$, $\zeta$ and $X_{\omega Y}$ can affect
significantly the neutron star properties.  The parameters of FTRMF model
are so calibrated that the quality of fit to finite nuclei, the properties
of nuclear matter at saturation density and hyperon-nucleon potentials
are almost the  same  for each of the parameterizations. Thus, these
parameterizations provide the right starting point to study the actual
variations in the properties of neutron star resulting from
the uncertainties in the EOS of dense matter.

The properties of non-rotating neutron stars  are obtained 
by integrating the  Tolman-Oppenheimer-Volkoff (TOV) equations
\cite{Weinberg72}. To solve the  TOV equations we use the EOS for the
matter consisting of nucleons, hyperons and leptons. The  composition
of matter at fixed total baryon density,
\begin{equation}
\rho = \sum_B\rho_B,
\label{eq:bd}
\end{equation}
are so determined that charge neutrality condition,
\begin{equation}
\sum_B q_{\scriptstyle B}\rho_{\scriptstyle B} +\sum_\ell q_\ell\rho_\ell = 0,
\label{eq:cn}
\end{equation}
and chemical equilibrium conditions,
\begin{eqnarray}
\label{eq:ce1}
\mu_{\scriptstyle B} = \mu_n-q_{\scriptstyle B}\mu_e\\
\mu_\mu = \mu_e
\label{eq:ce2}
\end{eqnarray}
are satisfied. In Eqs. (\ref{eq:cn}-\ref{eq:ce2}) $q$ and $\mu$
are the charge and chemical potential for various baryons and leptons
considered in our calculations.  For densities higher than $0.5\rho_0$,
the baryonic part of EOS is evaluated within the FTRMF model.
Whereas, the contributions of the electrons and muons to the EOS are
evaluated within  the Fermi gas approximation.  At densities lower
than 0.5$\rho_0$ down to $0.4\times 10^{-10}\rho_0$ we use the EOS of
Baym-Pethick-Sutherland (BPS) \cite{Baym71}.

In Fig. \ref{fig:eos} we plot the EOS for the pure neutron matter and
symmetric nuclear matter as a function  of number density for the selected
combinations of $\zeta$ and $\Delta r$. We see that the EOS for $\zeta
= 0.0$ is the most stiffest, and  as $\zeta$ increases the EOS becomes
softer. The softening of EOS with $\zeta$ is more pronounced at higher
densities.  In Fig. \ref{fig:mu} we plot our results for the neutron
and electron chemical potentials as a function of baryon density
obtained for the EOSs corresponding to the moderate values of $\Delta
r$ and $X_{\omega Y}$. The chemical potentials for other
particles can be evaluated using Eqs. (\ref{eq:ce1}) and (\ref{eq:ce2}).
The change in slope for neutron chemical potential vs. baryon density
is associated with appearance of  hyperons.  The decrease in $\mu_e$ for
$\rho\geqslant 2\rho_0$ is accompanied by the appearance of the $\Xi^-$
hyperons. The maximum values of $\mu_e$ is less than half of the bare
mass for kaons  which indicate that the presence of hyperons inhibits
the kaon condensation.

Let us  now consider various neutron star properties resulting from the
EOSs for the two different parameter sets referred hereafter as LY and
UY. These parameter sets are obtained using different combinations
of $\Delta r$, $\zeta$ and $X_{\omega Y}$. The parameters of LY set
are obtained with $\Delta r = 0.16\text{ fm}$, $\zeta = 0.06$ and
$X_{\omega Y} = 0.5$. Whereas, UY parameterization is obtained with
$\Delta r = 0.28\text{ fm}$, $\zeta = 0.0$ and $X_{\omega Y} = 0.8$.
Among all the parameterizations as obtained in Sec. \ref{sec:paramet},
LY and UY yield the softest and the stiffest EOSs, respectively. Thus,
maximum  variations in the neutron star properties can be studied using
the EOSs obtained for LY and UY parameter sets.  For the comparison,
we also present our results for the  L0 and U0 parameter sets similar to
LY and UY parameterizations, but, with no hyperons.  In Fig. \ref{fig:m-r}
we present our results for mass-radius relationship for LY, UY,
L0 and U0 parameterizations.  The region bounded by $R\leqslant
3GM/c^2$ is excluded by the causality limit \cite{Lattimer90}. The
line labeled by $\Delta I/I = 0.014$ is radius limit estimated by
Vela pulsar glitches \cite{Lattimer01}.  The rotation constraint as
indicated in Fig.  \ref{fig:m-r} is obtained using \cite{Lattimer04},
\begin{equation} \label{eq:kep} \nu_k = 1833\eta\left (\frac{\text
M}{\text M_\odot}\right )^{1/2}\left ( \frac{10 \text km}{R}\right
)^{3/2} \text Hz \end{equation} with $\eta = 0.57$ and $\nu_k =
1122$ Hz which corresponds to the frequency for the fastest rotating
neutron star present in the recently observed X-ray transient
XTE J1739-285 \cite{Kaaret06}.  The renormalization factor $\eta$
account for the effects due to deformation and gravity.  We also
calculate the variations in the radiation radius, \begin{equation}
R_\infty=\frac{R}{\sqrt{1-\frac{2GM}{Rc^2}}} \end{equation} for the
neutron star with the canonical mass $1.4M_\odot$.  It can be verified by
using the results for the LY and UY cases presented in Fig.  \ref{fig:m-r}
that $R_\infty$ lies in the range of $14.2 - 16.8\text{ km}$. Similarly,
without the inclusion of hyperons, the values of $R_\infty$ vary in the
range of $15.3 - 16.8\text{ km}$.  In Tables  IV  and V  we collect few
important bulk properties for the non-rotating neutron stars with maximum
and canonical masses.  We see that the  values of $M_{\text{max}}$ with
the inclusion of hyperons varies between $1.4 - 2.1M_\odot$.  Once the
contributions from the hyperons are ignored $M_{\text{max}}$ varies
between $1.7 - 2.4M_\odot$.  The values of  $R_{1.4}$ varies from $11.3
- 14.1\text{ km}$ and $12.5 - 14.1\text{ km}$ depending on whether the
hyperonic contributions are included or not.  Thus, combining our results
for the neutron stars  with and without hyperons we find that the values
of $M_{\text{ max}}$ and $R_{1.4}$ obtained within the FTRMF model can
vary about $1M_\odot$ and $3$ km, respectively.  These variations are
almost half of the ones obtained earlier  using FTRMF model in which
bulk nuclear observables and nuclear matter incompressibility were not
fitted appropriately.  The values of redshift given in the Tables IV
and V are obtained for the ratio $M/R$ as,

\begin{equation}
Z=\frac{1}{\sqrt{1-\frac{2GM}{Rc^2}}} - 1.
\label{eq:rshift}
\end{equation}
Our results for the  values of redshift for the neutron star with
canonical  mass are $0.22\pm 0.03$. It is also interesting to note that
$Z \geqslant 0.35$ only for the stars with masses $1.7M_\odot$ or larger.

In Fig. \ref{fig:thrd} we have plotted the threshold densities for various
hyperon species. In the same figure we also show the values of central
densities for the neutron stars with canonical mass and maximum mass.
The threshold density is lowest for the $\Lambda$ hyperons.  It is
interesting to note that  for the UY case the  threshold density for the
$\Lambda$ hyperons is almost equal to the  central density for the neutron
star with the canonical mass.  This implies that the properties of the
neutron star with the canonical mass do not get affected by the hyperons
for the UY parameterization. This is the reason that our results for the
mass and radius for U0 and UY parameterizations are very much similar
for the neutron stars with masses upto $1.6M_\odot$ as can be seen from
Fig. \ref{fig:m-r}.  The $\Sigma^+$ and $\Sigma^0$  hyperons do not appear
in density range relevant for the present study.  However, for the TM1
parameterization of the FTRMF model one finds that all kinds of hyperons
appear well below $7\rho_0$ \cite{Schaffner96,Shen02}.  This seems to
be due to large value of nuclear matter incompressibility coefficient
($K = 281$ MeV) associated with the TM1 parameter set.  In other words,
not only the variations in the properties of the neutron stars reduces
but also the chemical compositions for these  stars can become different
if the parameters of the FTRMF models are calibrated appropriately.
In Fig. \ref{fig:frac} we plot the particle fractions as a function of
radial coordinate.  These fractions are calculated for the neutron stars
with  $M_{\text{max}}$ =  $1.4M_\odot$ and $2.1M_\odot$  corresponding
to the LY (upper panel)  and UY (lower panel)  parameterizations,
respectively. The neutron fractions  in Fig.  \ref{fig:frac} are plotted
after dividing them by a factor of three. We see that the compositions
of the neutron stars shown in the upper and lower panels are not the
same. For the case of LY parameterizations, $\Xi^-$ and $\Sigma^-$
hyperons appear more or less simultaneously. For the UY case, $\Xi^0$
hyperons appear  instead of $\Sigma^-$ hyperons .  It is noteworthy
that for the case with UY parameterization the hyperons are the dominant
particles at the interior ($r < 4\text{ km}$) of the neutron star leading
to complete deleptonization.  We see from Fig. \ref{fig:frac} that the
proton fractions for both the cases  are greater than the critical value
($\sim 15\% $) for the Direct Urca process to occur \cite{Lattimer91}.

We now consider our results for neutron star properties at the canonical
and maximum masses for the set of EOSs obtained using all the different
parameterizations as given in Sec. \ref{sec:paramet}. These different
parameterizations correspond to the different combinations of the $\Delta
r$, $\zeta$ and $X_{\omega Y}$. The values of $\Delta r$, $\zeta$
and $X_{\omega Y}$ vary in the range $0.16 - 0.28\text{ fm}$, $0.0 -
0.06$ and $0.5 - 0.8$, respectively.  The knowledge of $M_{\text{max}}$
and ($R_{1.4}$) or moment of inertia (${\cal I}_{1.4}$) for neutron
star with canonical mass are very important in order to understand the
behaviour of the EOS over the wide range of density well above  $\rho_0$.
The discovery of the pulsars PSR J0737-3039A,B and PSR J0751+1807 have
raised hope for availability of more accurate information about these
quantities in near future.  The $M_{\text{max}}$ probes densest segment
of the EOS. Whereas, $R_{1.4}$ or ${\cal I}_{1.4}$ probes relatively
lower density region of EOS.  It is not possible to say a priori whether
or not $M_{\text{max}}$ is correlated to the properties of neutron star
with $1.4M_\odot$.  Earlier studies using FTRMF models indicate some
correlations between $M_{\text{max}}$ and $R_{1.4}$ \cite{Steiner05}.
Another study carried out for 25 EOSs taken from different models show
hardly any correlations between $M_{\text{max}}$ and ${\cal I}_{1.34}$
\cite{Bejger05}.  In Fig. \ref{fig:r14_z14} we plot the variations of
radius and the redshift for the neutron star with the canonical mass
as a function of $M_{\text{max}}$.  We see that $M_{\text{max}}$ varies
between $1.4 - 2.1M_\odot$ and $R_{1.4}$ varies between  $11.3 - 14.1$ km.
The vertical line at $M_{\text{max}} =1.6M_\odot$ corresponds to the mass
of the PSR J0751+1807 measured with $95\%$ confidence limit.  If  only
those EOSs are considered  for which $M_{\text{max}}\geqslant 1.6M_\odot$
then the value of  $R_{1.4}$ would  lie in the range of $12.8 - 14.1$ km.
This result is in reasonable agreement with $R_{1.4}=14.8^{+1.8}_{-1.6}$
km as deduced very recently by adequately fitting the high quality  X-ray
spectrum from the neutron star X7 in the globular cluster 47 Tucanae
\cite{Heinke06}.  We also note strong correlations of $M_{\text{max}}$
with $R_{1.4}$ and $Z_{1.4}$.  For a given value of $M_{\text{max}}$, the
spread in the values of $R_{1.4}$ is  $0.7\pm 0.1\text{ km}$. Only for
the $M_{\text{max}} \sim 1.4M_\odot$ we find that spread in the values
of $R_{1.4}$ is $\sim 0.3\text{ km}$. To understand it better, we list
in Table VI the values of $M_{\text{max}}$ and $R_{1.4}$ obtained for
the sets of  EOSs corresponding to the selected combinations of $\Delta
r$, $\zeta$ and $X_{\omega Y}$. For additional information we also give
in Table VI the values of $R_{\text{max}}$ which corresponds to the
radius of neutron star with maximum mass.  It is clear from the table
that for smaller $\zeta$ the value of $R_{1.4}$ varies with $\Delta r$
and is independent of $X_{\omega Y}$. This is due to the fact that for
smaller $\zeta$,  central density for neutron star with mass $1.4M_\odot$
is lower or almost equal to the threshold density for hyperons (see also
Fig. \ref{fig:thrd}). But, as $\zeta$ increases, the central density
becomes larger than the threshold densities for various hyperons,
thus, $R_{1.4}$ depends on $\Delta r$ as well as $X_{\omega Y}$.
In Fig. \ref{fig:rmax_zmax} we plot the variations of $R_{\text{max}}$
and $Z_{\text{max}}$ versus the maximum neutron star mass.  We see that
the correlations in the values of $M_{\text{max}}$ and $R_{\text{max}}$
are stronger than the ones observed in the case of $M_{\text{max}}$ and
$R_{1.4}$. The spread in the values of $R_{\text{max}}$ is only $0.2\pm
0.1\text{ km}$ for a fixed value of $M_{\text{max}}$.  The values of
$R_{\text{max}}$ do not depend  strongly on the choice of $\Delta r$
as can be seen from Table VI.  The horizontal line in the lower panel
corresponds to the measured value of the redshift, $Z = 0.35$, for
the neutron star EXO 0748-676 \cite{Ozel06}.  For $Z=0.35$, we find
that the $M_{\text{max}}$ is $\sim 1.8M_\odot$ and the corresponding
radius is $\sim 12\text{ km}$. These values for neutron star masses
and the corresponding radii are in reasonable agreement with the best
suggested value of the mass $1.8M_\odot$ and radius $11.5\text{ km} $
corresponding to Z=0.35 \cite{Villarreal04}.

To this end, we would like to mention  that the calculations are repeated
for an attractive $\Sigma - N$ potentials by assuming $U_{\Sigma}^{(N)}
= -30 $ MeV in Eq. (\ref{eq:uyn}).  We find that  with this choice of
$U_\Sigma^{(N)}$ our results for the variations in the $M_{\text{max}}$
and $R_{1.4}$ do not get affected. However, the threshold density for
$\Sigma^-$ hyperon becomes lowest and $\Xi^-$ hyperon does not appear
even for the maximum  neutron star mass.  It must be pointed out that
the tensor coupling of $\omega$-meson to the hyperons, not considered
in the present work, could increase the value of $M_{\text{max}}$ by
about $0.1M_\odot$ \cite{Sugahara94a}.  We also remark that the effects
due to the exchange and the correlations are not considered explicitly.
But,they are taken into account at least partly through the non-linear
self and mixed interactions of the mesons \cite{Furnstahl96,Furnstahl97}.
The Eqs. (\ref{eq:sigma}) - (\ref{eq:photon}) can be interpreted as
the Khon-Sham equations in relativistic case and in this sense they
include effects beyond the Hartree approach through the non-linear
couplings. However, a more accurate treatment of the exchange and
correlation effects should be pursued \cite{Greco01,Panda06}.

\section{Rotating neutron stars}\label{sec:rot}

The properties of neutron stars can get significantly affected in
the presence of rotation. The effects of rotation on the neutron star
properties are pronounced when the frequency of rotation is close to its
Keplerian limit. Earlier studies indicate that  the Keplerian frequency
is $\sim 1000$ Hz for the neutron stars with mass around $1M_\odot$
\cite{Shapiro83}. Only very recently \cite{Kaaret06}, a neutron star
rotating at 1122 Hz is  discovered in the X-ray transient XTE J1739-285. In
this section we shall discuss our results for the rotating neutron
stars obtained using the extended FTRMF model.  These results are obtained
by solving the Einstein equations for stationary axi-symmetric spacetime.
The numerical computations are performed using the code written by
Stergioulas \cite{Stergioulas95}.

In Fig. \ref{fig:kep_seq} we plot the neutron star mass versus
the circumferential equatorial radius $R_{\text{eq}}$ for the Keplerian
sequences obtained using EOSs for the  LY, UY, L0 and U0 parameterizations
of our model.  The maximum mass of the neutron stars vary between $1.7 -
2.5M_\odot$ and  $2.0 - 3.0M_\odot$ for the cases with and without the
hyperons respectively. The values of $R^{1.4}_{\text {eq}}$
lie in the range of $18.4 - 20.0$ km irrespective of whether or
not hyperonic degrees of freedom are included. Because, the central
density for the canonical mass in the presence of rotation becomes
lower than the threshold densities for the hyperons.  The Keplerian
frequencies at maximum  neutron star mass for various cases shown in
Fig. \ref{fig:kep_seq}  lie in the range of $1320 - 1560$ Hz . This
means that all the EOSs obtained in the present work can yield
neutron stars rotating at 1122 Hz.  In Fig. \ref{fig:mreq_1122}
we plot the mass and the corresponding values for $R_{\text{eq}}$
for the neutron star rotating at $1122$ Hz. The lower and upper
bounds on the radii $R_{\text{eq}}$ are determined by the setting-in
of the axi-symmetric perturbation and mass-shedding instabilities,
respectively \cite{Bejger06}.  The maximum values of $R_{\text{eq}}$
are well fitted by \cite{Bejger06} 
\begin{equation} \label{eq:rmax}
R_{\text {eq}}^{\text{max}}=13.87\left (\frac{M}{M_\odot}\right )^{1/3} \,
\text {km} 
\end{equation}
which can be obtained using $\nu_k = 1122$ Hz and $\eta =1$ in Eq.
(\ref{eq:kep}).  In Table \ref{tab:ns_1122}, we give the minium and
maximum values for the $R_{\text{eq}}$ and the corresponding neutron
star mass for the various cases plotted in Fig. \ref{fig:mreq_1122}.
We get $R_{\text {eq}}^{\text{min}} = 12.1 - 13.8$ km and $R_{\text
{eq}}^{\text{max}} = 16.5 - 18.7$ km.  The values of $M(R_{\text
{eq}}^{\text{min}})$ and $M(R_{\text {eq}}^{\text{max}})$ are in the
range of $1.6 - 2.7 M_\odot$ and $1.7 - 2.6M_\odot$, respectively.
The absolute difference between the $M(R_{\text {eq}}^{\text{min}})$
and $M(R_{\text {eq}}^{\text{max}})$ which gives the variations in the
neutron star mass for a given EOS is at most $0.2M_\odot$.  We also
find that the baryonic mass for the neutron stars rotating with 1122 Hz
for all the cases considered here are larger than the maximum baryonic
mass for the corresponding non-rotating sequences. This suggests that
the recently discovered fastest rotating neutron star rotating with
1122 Hz is supra massive.  We also list in Table \ref{tab:ns_1122} the
values for the $r_{\text{pole}}/r_{\text{eq}}$ known as the flattening
parameter and $T/\mid W\mid$ where $T$ the kinetic energy and $W$ the
gravitational energy.

\section{Conclusions}
\label{sec:conc}

We have used the extended FTRMF model to obtain a new set of EOSs with
the different high density behaviour. These EOSs are then employed to
study the variations in the properties of non-rotating and rotating
neutron stars.  The high density behaviour of the EOS  which is not yet
well constrained is varied by choosing the different values of the
$\omega$-meson self-coupling and the couplings of $\omega$-meson to
the various hyperons in our model.  The different values for these
couplings are so chosen that they span the entire range as often
considered in the earlier  works.  The remaining parameters of the models
are calibrated to yield reasonable fit to the bulk nuclear observables,
nuclear matter incompressibility coefficient and hyperon-nucleon potential
depths.  The properties of finite nuclei and nuclear matter associated with
each of the parameterizations used for obtaining EOSs  can be summarized
as follows. The rms errors for the total binding energies and charge
radii calculated for the nuclei considered in the fits are $1.5 - 1.8$
MeV and $0.025 - 0.040$ fm.  The binding energy per nucleon is
$16.11\pm 0.04$ MeV, saturation density is $0.148\pm 0.003$ fm$^{-3}$
and nuclear matter incompressibility coefficient is $230.24\pm 9.80$ MeV.

The values of $M_{\text{max}}$ for the non-rotating neutron stars
composed of nucleons and  hyperons in $\beta$ equilibrium can vary
between $1.4 - 2.1M_\odot$. The radius $R_{1.4}$ for neutron star
can vary in the range of $11.3 - 14.1$ km.  The values of $R_{1.4}$
narrow down to only $12.8 - 14.1$ km if one considers the EOSs for which
$M_{\text{max}}$ is larger than $1.6M_\odot$ as it is the highest mass
measured for PSR J0751+1807 with $95\%$ confidence limit.  This result is
in reasonable agreement with $R_{1.4}=14.8^{+1.8}_{-1.6}$ km as deduced
very recently by adequately fitting the high quality  X-ray spectrum from
the neutron star X7 in the globular cluster 47 Tucanae \cite{Heinke06}.
We also note strong correlations between the values of $M_{\text{max}}$
and $R_{1.4}$.  The values of the redshift for the neutron stars with
canonical and maximum masses are also calculated. The redshift for the
neutron star with canonical mass obtained for different EOSs varies in
between $0.19 - 0.25$.   The maximum value for the redshift is 0.41 which
corresponds to the maximum neutron star mass of $2.1M_\odot$. For the
measured value of redshift equal to 0.35, we find that the neutron star
mass is $\sim 1.8M_\odot$ and the corresponding radius is $\sim 12\text{
km}$.  These values for neutron star masses and the corresponding radii
are in reasonable agreement with the best suggested value of the mass
$1.8M_\odot$ and radius $11.5\text{ km} $ for Z=0.35 \cite{Villarreal04}.
For the sake of comparison we have presented our results  obtained without
the inclusions of the hyperons.  In this case the $M_{\text{max}}$
and $R_{1.4}$ lie in the range of $1.7 - 2.4M_\odot$ and $12.5 -
14.1$ km, respectively.

We use our EOSs to compute the properties of the rotating neutron
stars. In particular, we studied the mass and the circumferential
equatorial radius for the neutron star rotating at 1122 Hz as recently
observed \cite{Kaaret06}.  Our results for different EOSs indicate
that the mass for such a star can lie within $1.6 - 2.7M_\odot$. The
minimum values for the circumferential equatorial radius determined
by the onset of the instability with respect to the axi-symmetric
perturbation are found to vary in the range of $12.1 - 13.8$ km.
The maximum values for the circumferential equatorial radius obtained by
the mass-shedding limit vary within $16.5 - 18.7$ km. Looking into the
results for the baryonic mass we find that the neutron star rotating at
1122 Hz are supra massive for our EOSs.

\begin{acknowledgments}
We would like to thank D. Bandyopadhyay for useful discussions. This work
was supported in part by the University Grant Commission under grant \#
F.17-40/98 (SA-I).  Shashi K. Dhiman and Raj Kumar greatly acknowledge the
financial support from Saha Institute of Nuclear Physics, Kolkata, India.

\end{acknowledgments}

\newpage

\begin{thebibliography}{61}
\expandafter\ifx\csname natexlab\endcsname\relax\def\natexlab#1{#1}\fi
\expandafter\ifx\csname bibnamefont\endcsname\relax
  \def\bibnamefont#1{#1}\fi
\expandafter\ifx\csname bibfnamefont\endcsname\relax
  \def\bibfnamefont#1{#1}\fi
\expandafter\ifx\csname citenamefont\endcsname\relax
  \def\citenamefont#1{#1}\fi
\expandafter\ifx\csname url\endcsname\relax
  \def\url#1{\texttt{#1}}\fi
\expandafter\ifx\csname urlprefix\endcsname\relax\def\urlprefix{URL }\fi
\providecommand{\bibinfo}[2]{#2}
\providecommand{\eprint}[2][]{\url{#2}}

\bibitem[{\citenamefont{Thorsett and Chakrabarty}(1999)}]{Thorsett99}
\bibinfo{author}{\bibfnamefont{S.~E.} \bibnamefont{Thorsett}} \bibnamefont{and}
  \bibinfo{author}{\bibfnamefont{D.}~\bibnamefont{Chakrabarty}},
  \bibinfo{journal}{Astrophys. J} \textbf{\bibinfo{volume}{512}},
  \bibinfo{pages}{288} (\bibinfo{year}{1999}).

\bibitem[{\citenamefont{Burgay et~al.}(2003)\citenamefont{Burgay, D'Amico,
  Possenti, Manchester, Lyne, Joshi, McLaughlin, M.~Kramer, Kalogera, Kim
  et~al.}}]{Burgay03}
\bibinfo{author}{\bibfnamefont{M.}~\bibnamefont{Burgay}},
  \bibinfo{author}{\bibfnamefont{N.}~\bibnamefont{D'Amico}},
  \bibinfo{author}{\bibfnamefont{A.}~\bibnamefont{Possenti}},
  \bibinfo{author}{\bibfnamefont{R.~N.} \bibnamefont{Manchester}},
  \bibinfo{author}{\bibfnamefont{A.~G.} \bibnamefont{Lyne}},
  \bibinfo{author}{\bibfnamefont{B.~C.} \bibnamefont{Joshi}},
  \bibinfo{author}{\bibfnamefont{M.~A.} \bibnamefont{McLaughlin}},
  \bibinfo{author}{\bibfnamefont{F.~C.} \bibnamefont{M.~Kramer},
  \bibfnamefont{J.~M.~Sarkissian}},
  \bibinfo{author}{\bibfnamefont{V.}~\bibnamefont{Kalogera}},
  \bibinfo{author}{\bibfnamefont{C.}~\bibnamefont{Kim}}, \bibnamefont{et~al.},
  \bibinfo{journal}{Nature} \textbf{\bibinfo{volume}{426}},
  \bibinfo{pages}{531} (\bibinfo{year}{2003}).

\bibitem[{\citenamefont{Lyne and et~al}(2004)}]{Lyne04}
\bibinfo{author}{\bibfnamefont{A.}~\bibnamefont{Lyne}} \bibnamefont{}
  \bibinfo{author}{\bibnamefont{{\it et~al}}}, \bibinfo{journal}{Science}
  \textbf{\bibinfo{volume}{303}}, \bibinfo{pages}{1153} (\bibinfo{year}{2004}).

\bibitem[{\citenamefont{Nice et~al.}(2005)\citenamefont{Nice, Splaver, Stairs,
  Lohmer, Jessner, Kramer, and Cordes}}]{Nice05a}
\bibinfo{author}{\bibfnamefont{D.~J.} \bibnamefont{Nice}},
  \bibinfo{author}{\bibfnamefont{E.}~\bibnamefont{Splaver}},
  \bibinfo{author}{\bibfnamefont{I.}~\bibnamefont{Stairs}},
  \bibinfo{author}{\bibfnamefont{O.}~\bibnamefont{Lohmer}},
  \bibinfo{author}{\bibfnamefont{A.~J.~A.} \bibnamefont{Jessner}},
  \bibinfo{author}{\bibfnamefont{M.}~\bibnamefont{Kramer}}, \bibnamefont{and}
  \bibinfo{author}{\bibfnamefont{J.}~\bibnamefont{Cordes}},
  \bibinfo{journal}{Astrophys. J.} \textbf{\bibinfo{volume}{634}},
  \bibinfo{pages}{1242} (\bibinfo{year}{2005}).

\bibitem[{\citenamefont{Rutledge
  et~al.}(2002{\natexlab{a}})\citenamefont{Rutledge, Bildsten, Brown, Pavlov,
  and Zavlin}}]{Rutledge02}
\bibinfo{author}{\bibfnamefont{R.}~\bibnamefont{Rutledge}},
  \bibinfo{author}{\bibfnamefont{L.}~\bibnamefont{Bildsten}},
  \bibinfo{author}{\bibfnamefont{E.}~\bibnamefont{Brown}},
  \bibinfo{author}{\bibfnamefont{G.}~\bibnamefont{Pavlov}}, \bibnamefont{and}
  \bibinfo{author}{\bibfnamefont{V.}~\bibnamefont{Zavlin}},
  \bibinfo{journal}{Astrophys. J.} \textbf{\bibinfo{volume}{578}},
  \bibinfo{pages}{405} (\bibinfo{year}{2002}{\natexlab{a}}).

\bibitem[{\citenamefont{Rutledge
  et~al.}(2002{\natexlab{b}})\citenamefont{Rutledge, Bildsten, Brown, Pavlov,
  and Zavlin}}]{Rutledge02a}
\bibinfo{author}{\bibfnamefont{R.}~\bibnamefont{Rutledge}},
  \bibinfo{author}{\bibfnamefont{L.}~\bibnamefont{Bildsten}},
  \bibinfo{author}{\bibfnamefont{E.}~\bibnamefont{Brown}},
  \bibinfo{author}{\bibfnamefont{G.}~\bibnamefont{Pavlov}}, \bibnamefont{and}
  \bibinfo{author}{\bibfnamefont{V.}~\bibnamefont{Zavlin}},
  \bibinfo{journal}{Astrophys. J.} \textbf{\bibinfo{volume}{577}},
  \bibinfo{pages}{346} (\bibinfo{year}{2002}{\natexlab{b}}).

\bibitem[{\citenamefont{Gendre et~al.}(2003)\citenamefont{Gendre, Barret, and
  Webb}}]{Gendre03}
\bibinfo{author}{\bibfnamefont{B.}~\bibnamefont{Gendre}},
  \bibinfo{author}{\bibfnamefont{D.}~\bibnamefont{Barret}}, \bibnamefont{and}
  \bibinfo{author}{\bibfnamefont{N.~A.} \bibnamefont{Webb}},
  \bibinfo{journal}{Astrono. Astrophys.} \textbf{\bibinfo{volume}{400}},
  \bibinfo{pages}{521} (\bibinfo{year}{2003}).

\bibitem[{\citenamefont{Becker and et~al.}(2003)}]{Becker03}
\bibinfo{author}{\bibfnamefont{W.}~\bibnamefont{Becker}} \bibnamefont{and}
  \bibinfo{author}{\bibnamefont{et~al.}}, \bibinfo{journal}{Astrophys. J.}
  \textbf{\bibinfo{volume}{594}}, \bibinfo{pages}{364} (\bibinfo{year}{2003}).

\bibitem[{\citenamefont{Cottam et~al.}(2002)\citenamefont{Cottam, Paerels, and
  Mendez}}]{Cottam02}
\bibinfo{author}{\bibfnamefont{J.}~\bibnamefont{Cottam}},
  \bibinfo{author}{\bibfnamefont{F.}~\bibnamefont{Paerels}}, \bibnamefont{and}
  \bibinfo{author}{\bibfnamefont{M.}~\bibnamefont{Mendez}},
  \bibinfo{journal}{Nature} \textbf{\bibinfo{volume}{420}}, \bibinfo{pages}{51}
  (\bibinfo{year}{2002}).

\bibitem[{\citenamefont{Kaaret et~al.}(2006)\citenamefont{Kaaret, Prieskorn,
  in~'t Zand, Brandt, Lund, Mereghetti, Gotz, Kuulkers, and
  Tomsick}}]{Kaaret06}
\bibinfo{author}{\bibfnamefont{P.}~\bibnamefont{Kaaret}},
  \bibinfo{author}{\bibfnamefont{Z.}~\bibnamefont{Prieskorn}},
  \bibinfo{author}{\bibfnamefont{J.}~\bibnamefont{in~'t Zand}},
  \bibinfo{author}{\bibfnamefont{S.}~\bibnamefont{Brandt}},
  \bibinfo{author}{\bibfnamefont{N.}~\bibnamefont{Lund}},
  \bibinfo{author}{\bibfnamefont{S.}~\bibnamefont{Mereghetti}},
  \bibinfo{author}{\bibfnamefont{D.}~\bibnamefont{Gotz}},
  \bibinfo{author}{\bibfnamefont{E.}~\bibnamefont{Kuulkers}}, \bibnamefont{and}
  \bibinfo{author}{\bibfnamefont{J.}~\bibnamefont{Tomsick}},
  \bibinfo{journal}{Astrophys. J.} \textbf{\bibinfo{volume}{657}},
  \bibinfo{pages}{L97} (\bibinfo{year}{2006}).

\bibitem[{\citenamefont{Lavagetto et~al.}(2006)\citenamefont{Lavagetto,
  Bombaci, D'Ai', Vidana, and Robba}}]{Lavagetto06}
\bibinfo{author}{\bibfnamefont{G.}~\bibnamefont{Lavagetto}},
  \bibinfo{author}{\bibfnamefont{I.}~\bibnamefont{Bombaci}},
  \bibinfo{author}{\bibfnamefont{A.}~\bibnamefont{D'Ai'}},
  \bibinfo{author}{\bibfnamefont{I.}~\bibnamefont{Vidana}}, \bibnamefont{and}
  \bibinfo{author}{\bibfnamefont{N.}~\bibnamefont{Robba}},
  \bibinfo{journal}{astro-ph/0612061}  (\bibinfo{year}{2006}).

\bibitem[{\citenamefont{Pandharipande and Smith}(1975)}]{Pandharipande75}
\bibinfo{author}{\bibfnamefont{V.~R.} \bibnamefont{Pandharipande}}
  \bibnamefont{and} \bibinfo{author}{\bibfnamefont{R.~A.} \bibnamefont{Smith}},
  \bibinfo{journal}{Nucl. Phys.} \textbf{\bibinfo{volume}{A237}},
  \bibinfo{pages}{507} (\bibinfo{year}{1975}).

\bibitem[{\citenamefont{Chabanat et~al.}(1997)\citenamefont{Chabanat, Bonche,
  Haensel, Meyer, and Schaeffer}}]{Chabanat97}
\bibinfo{author}{\bibfnamefont{E.}~\bibnamefont{Chabanat}},
  \bibinfo{author}{\bibfnamefont{P.}~\bibnamefont{Bonche}},
  \bibinfo{author}{\bibfnamefont{P.}~\bibnamefont{Haensel}},
  \bibinfo{author}{\bibfnamefont{J.}~\bibnamefont{Meyer}}, \bibnamefont{and}
  \bibinfo{author}{\bibfnamefont{R.}~\bibnamefont{Schaeffer}},
  \bibinfo{journal}{Nucl. Phys.} \textbf{\bibinfo{volume}{A627}},
  \bibinfo{pages}{710} (\bibinfo{year}{1997}).

\bibitem[{\citenamefont{Stone et~al.}(2003)\citenamefont{Stone, Miller,
  Koncewicz, Stevenson, and Strayer}}]{Stone03}
\bibinfo{author}{\bibfnamefont{J.~R.} \bibnamefont{Stone}},
  \bibinfo{author}{\bibfnamefont{J.~C.} \bibnamefont{Miller}},
  \bibinfo{author}{\bibfnamefont{R.}~\bibnamefont{Koncewicz}},
  \bibinfo{author}{\bibfnamefont{P.~D.} \bibnamefont{Stevenson}},
  \bibnamefont{and} \bibinfo{author}{\bibfnamefont{M.~R.}
  \bibnamefont{Strayer}}, \bibinfo{journal}{Phys. Rev. C}
  \textbf{\bibinfo{volume}{68}}, \bibinfo{pages}{034324}
  (\bibinfo{year}{2003}).

\bibitem[{\citenamefont{Monras}(2005)}]{Mornas05}
\bibinfo{author}{\bibfnamefont{L.}~\bibnamefont{Monras}},
  \bibinfo{journal}{Eur. Phys. J.} \textbf{\bibinfo{volume}{A24}},
  \bibinfo{pages}{293} (\bibinfo{year}{2005}).

\bibitem[{\citenamefont{Agrawal et~al.}(2006)\citenamefont{Agrawal, Dhiman, and
  Kumar}}]{Agrawal06}
\bibinfo{author}{\bibfnamefont{B.~K.} \bibnamefont{Agrawal}},
  \bibinfo{author}{\bibfnamefont{S.~K.} \bibnamefont{Dhiman}},
  \bibnamefont{and} \bibinfo{author}{\bibfnamefont{R.}~\bibnamefont{Kumar}},
  \bibinfo{journal}{Phys. Rev. C} \textbf{\bibinfo{volume}{73}},
  \bibinfo{pages}{034319} (\bibinfo{year}{2006}).

\bibitem[{\citenamefont{Prakash et~al.}(1995)\citenamefont{Prakash, Cooke, and
  Lattimer}}]{Prakash95}
\bibinfo{author}{\bibfnamefont{M.}~\bibnamefont{Prakash}},
  \bibinfo{author}{\bibfnamefont{J.~R.} \bibnamefont{Cooke}}, \bibnamefont{and}
  \bibinfo{author}{\bibfnamefont{J.~M.} \bibnamefont{Lattimer}},
  \bibinfo{journal}{Phys. Rev. D} \textbf{\bibinfo{volume}{52}},
  \bibinfo{pages}{661} (\bibinfo{year}{1995}).

\bibitem[{\citenamefont{Glendenning and
  Schaffner-Bielich}(1999)}]{Glendenning99}
\bibinfo{author}{\bibfnamefont{N.~K.} \bibnamefont{Glendenning}}
  \bibnamefont{and}
  \bibinfo{author}{\bibfnamefont{J.}~\bibnamefont{Schaffner-Bielich}},
  \bibinfo{journal}{Phys. Rev. C} \textbf{\bibinfo{volume}{60}},
  \bibinfo{pages}{025803} (\bibinfo{year}{1999}).

\bibitem[{\citenamefont{Steiner et~al.}(2005)\citenamefont{Steiner, Prakash,
  Lattimer, and Ellis}}]{Steiner05}
\bibinfo{author}{\bibfnamefont{A.~W.} \bibnamefont{Steiner}},
  \bibinfo{author}{\bibfnamefont{M.}~\bibnamefont{Prakash}},
  \bibinfo{author}{\bibfnamefont{J.~M.} \bibnamefont{Lattimer}},
  \bibnamefont{and} \bibinfo{author}{\bibfnamefont{P.}~\bibnamefont{Ellis}},
  \bibinfo{journal}{Phys. Rep.} \textbf{\bibinfo{volume}{411}},
  \bibinfo{pages}{325} (\bibinfo{year}{2005}).

\bibitem[{\citenamefont{M{\"u}ther et~al.}(1987)\citenamefont{M{\"u}ther,
  Prakash, and Ainsworth}}]{Muther87}
\bibinfo{author}{\bibfnamefont{H.}~\bibnamefont{M{\"u}ther}},
  \bibinfo{author}{\bibfnamefont{M.}~\bibnamefont{Prakash}}, \bibnamefont{and}
  \bibinfo{author}{\bibfnamefont{T.~L.} \bibnamefont{Ainsworth}},
  \bibinfo{journal}{Phys. Lett.} \textbf{\bibinfo{volume}{B199}},
  \bibinfo{pages}{469} (\bibinfo{year}{1987}).

\bibitem[{\citenamefont{Engvik et~al.}(1994)\citenamefont{Engvik,
  Hjorth-Jensen, Osnes, Bao, and \O{}stgaard}}]{Engvik94}
\bibinfo{author}{\bibfnamefont{L.}~\bibnamefont{Engvik}},
  \bibinfo{author}{\bibfnamefont{M.}~\bibnamefont{Hjorth-Jensen}},
  \bibinfo{author}{\bibfnamefont{E.}~\bibnamefont{Osnes}},
  \bibinfo{author}{\bibfnamefont{G.}~\bibnamefont{Bao}}, \bibnamefont{and}
  \bibinfo{author}{\bibfnamefont{E.}~\bibnamefont{\O{}stgaard}},
  \bibinfo{journal}{Phys. Rev. Lett.} \textbf{\bibinfo{volume}{73}},
  \bibinfo{pages}{2650} (\bibinfo{year}{1994}).

\bibitem[{\citenamefont{Engvik et~al.}(1996)\citenamefont{Engvik, Osnes,
  Hjorth-Jensen, Bao, and \O{}stgaard}}]{Engvik96}
\bibinfo{author}{\bibfnamefont{L.}~\bibnamefont{Engvik}},
  \bibinfo{author}{\bibfnamefont{E.}~\bibnamefont{Osnes}},
  \bibinfo{author}{\bibfnamefont{M.}~\bibnamefont{Hjorth-Jensen}},
  \bibinfo{author}{\bibfnamefont{G.}~\bibnamefont{Bao}}, \bibnamefont{and}
  \bibinfo{author}{\bibfnamefont{E.}~\bibnamefont{\O{}stgaard}},
  \bibinfo{journal}{Astrophys. J.} \textbf{\bibinfo{volume}{469}},
  \bibinfo{pages}{794} (\bibinfo{year}{1996}).

\bibitem[{\citenamefont{Schulze et~al.}(2006)\citenamefont{Schulze, Polls,
  Ramos, and Vidana}}]{Schulze06}
\bibinfo{author}{\bibfnamefont{H.~J.} \bibnamefont{Schulze}},
  \bibinfo{author}{\bibfnamefont{A.}~\bibnamefont{Polls}},
  \bibinfo{author}{\bibfnamefont{A.}~\bibnamefont{Ramos}}, \bibnamefont{and}
  \bibinfo{author}{\bibfnamefont{I.}~\bibnamefont{Vidana}},
  \bibinfo{journal}{Phys. Rev. C} \textbf{\bibinfo{volume}{73}},
  \bibinfo{pages}{058801} (\bibinfo{year}{2006}).

\bibitem[{\citenamefont{M{\"u}ller and Serot}(1996)}]{Muller96}
\bibinfo{author}{\bibfnamefont{H.}~\bibnamefont{M{\"u}ller}} \bibnamefont{and}
  \bibinfo{author}{\bibfnamefont{B.~D.} \bibnamefont{Serot}},
  \bibinfo{journal}{Nucl. Phys.} \textbf{\bibinfo{volume}{A606}},
  \bibinfo{pages}{508} (\bibinfo{year}{1996}).

\bibitem[{\citenamefont{Taurines et~al.}(2001)\citenamefont{Taurines,
  Vasconcellos, Malherio, and Chiapparini}}]{Taurines01}
\bibinfo{author}{\bibfnamefont{A.~R.} \bibnamefont{Taurines}},
  \bibinfo{author}{\bibfnamefont{C.~A.~Z.} \bibnamefont{Vasconcellos}},
  \bibinfo{author}{\bibfnamefont{M.}~\bibnamefont{Malherio}}, \bibnamefont{and}
  \bibinfo{author}{\bibfnamefont{M.}~\bibnamefont{Chiapparini}},
  \bibinfo{journal}{Phys. Rev. C} \textbf{\bibinfo{volume}{63}},
  \bibinfo{pages}{065801} (\bibinfo{year}{2001}).

\bibitem[{\citenamefont{Jha et~al.}(2006)\citenamefont{Jha, Raina, Panda, and
  Patra}}]{Jha06}
\bibinfo{author}{\bibfnamefont{T.~K.} \bibnamefont{Jha}},
  \bibinfo{author}{\bibfnamefont{P.~K.} \bibnamefont{Raina}},
  \bibinfo{author}{\bibfnamefont{P.~K.} \bibnamefont{Panda}}, \bibnamefont{and}
  \bibinfo{author}{\bibfnamefont{S.~K.} \bibnamefont{Patra}},
  \bibinfo{journal}{Phys. Rev. C} \textbf{\bibinfo{volume}{74}},
  \bibinfo{pages}{055803} (\bibinfo{year}{2006}).

\bibitem[{\citenamefont{Reinhard}(1999)}]{Reinhard99}
\bibinfo{author}{\bibfnamefont{P.~G.} \bibnamefont{Reinhard}},
  \bibinfo{journal}{Nucl. Phys.} \textbf{\bibinfo{volume}{A649}},
  \bibinfo{pages}{305c} (\bibinfo{year}{1999}).

\bibitem[{\citenamefont{Youngblood et~al.}(2002)\citenamefont{Youngblood, Lui,
  and Clark}}]{Youngblood02}
\bibinfo{author}{\bibfnamefont{D.~H.} \bibnamefont{Youngblood}},
  \bibinfo{author}{\bibfnamefont{Y.-W.} \bibnamefont{Lui}}, \bibnamefont{and}
  \bibinfo{author}{\bibfnamefont{H.~L.} \bibnamefont{Clark}},
  \bibinfo{journal}{Phys. Rev. C} \textbf{\bibinfo{volume}{65}},
  \bibinfo{pages}{034302} (\bibinfo{year}{2002}).

\bibitem[{\citenamefont{Furnstahl et~al.}(1996)\citenamefont{Furnstahl, Serot,
  and Tang}}]{Furnstahl96}
\bibinfo{author}{\bibfnamefont{R.}~\bibnamefont{Furnstahl}},
  \bibinfo{author}{\bibfnamefont{B.~D.} \bibnamefont{Serot}}, \bibnamefont{and}
  \bibinfo{author}{\bibfnamefont{H.-B.} \bibnamefont{Tang}},
  \bibinfo{journal}{Nucl. Phys.} \textbf{\bibinfo{volume}{A598}},
  \bibinfo{pages}{539} (\bibinfo{year}{1996}).

\bibitem[{\citenamefont{Furnstahl et~al.}(1997)\citenamefont{Furnstahl, Serot,
  and Tang}}]{Furnstahl97}
\bibinfo{author}{\bibfnamefont{R.}~\bibnamefont{Furnstahl}},
  \bibinfo{author}{\bibfnamefont{B.~D.} \bibnamefont{Serot}}, \bibnamefont{and}
  \bibinfo{author}{\bibfnamefont{H.-B.} \bibnamefont{Tang}},
  \bibinfo{journal}{Nucl. Phys.} \textbf{\bibinfo{volume}{A615}},
  \bibinfo{pages}{441} (\bibinfo{year}{1997}).
\bibitem[{\citenamefont{Serot and Walecka}(1997)}]{Serot97}
\bibinfo{author}{\bibfnamefont{B.~D.} \bibnamefont{Serot}} \bibnamefont{and}
  \bibinfo{author}{\bibfnamefont{J.~D.} \bibnamefont{Walecka}},
  \bibinfo{journal}{Int. J. Mod. Phys. E} \textbf{\bibinfo{volume}{6}},
  \bibinfo{pages}{515} (\bibinfo{year}{1997}).

\bibitem[{\citenamefont{Horowitz and Piekarewicz}(2001)}]{Horowitz01}
\bibinfo{author}{\bibfnamefont{C.~J.} \bibnamefont{Horowitz}} \bibnamefont{and}
  \bibinfo{author}{\bibfnamefont{J.}~\bibnamefont{Piekarewicz}},
  \bibinfo{journal}{Phys. Rev. Lett.} \textbf{\bibinfo{volume}{86}},
  \bibinfo{pages}{5647} (\bibinfo{year}{2001}).

\bibitem[{\citenamefont{Furnstahl}(2002)}]{Furnstahl02}
\bibinfo{author}{\bibfnamefont{R.}~\bibnamefont{Furnstahl}},
  \bibinfo{journal}{Nucl. Phys.} \textbf{\bibinfo{volume}{A706}},
  \bibinfo{pages}{85} (\bibinfo{year}{2002}).

\bibitem[{\citenamefont{Sil et~al.}(2005)\citenamefont{Sil, Centelles, Vinas,
  and Piekarewicz}}]{Sil05}
\bibinfo{author}{\bibfnamefont{T.}~\bibnamefont{Sil}},
  \bibinfo{author}{\bibfnamefont{M.}~\bibnamefont{Centelles}},
  \bibinfo{author}{\bibfnamefont{X.}~\bibnamefont{Vinas}}, \bibnamefont{and}
  \bibinfo{author}{\bibfnamefont{J.}~\bibnamefont{Piekarewicz}},
  \bibinfo{journal}{Phys. Rev.} \textbf{\bibinfo{volume}{C71}},
  \bibinfo{pages}{045502} (\bibinfo{year}{2005}).

\bibitem[{\citenamefont{Lattimer et~al.}(1991)\citenamefont{Lattimer, Pethick,
  M.Prakash, and Haensel}}]{Lattimer91}
\bibinfo{author}{\bibfnamefont{J.~M.} \bibnamefont{Lattimer}},
  \bibinfo{author}{\bibfnamefont{C.~J.} \bibnamefont{Pethick}},
  \bibinfo{author}{\bibnamefont{M.Prakash}}, \bibnamefont{and}
  \bibinfo{author}{\bibfnamefont{P.}~\bibnamefont{Haensel}},
  \bibinfo{journal}{Phys. Rev. Lett.} \textbf{\bibinfo{volume}{66}},
  \bibinfo{pages}{2701} (\bibinfo{year}{1991}).

\bibitem[{\citenamefont{Y.Sugahara and
  H.Toki}(1994{\natexlab{a}})}]{Sugahara94}
\bibinfo{author}{\bibnamefont{Y.Sugahara}} \bibnamefont{and}
  \bibinfo{author}{\bibnamefont{H.Toki}}, \bibinfo{journal}{Nucl. Phys.}
  \textbf{\bibinfo{volume}{A579}}, \bibinfo{pages}{557}
  (\bibinfo{year}{1994}{\natexlab{a}}).

\bibitem[{\citenamefont{Chen et~al.}(2005)\citenamefont{Chen, Ko, and
  Li}}]{Chen05}
\bibinfo{author}{\bibfnamefont{L.-W.} \bibnamefont{Chen}},
  \bibinfo{author}{\bibfnamefont{C.~M.} \bibnamefont{Ko}}, \bibnamefont{and}
  \bibinfo{author}{\bibfnamefont{B.-A.} \bibnamefont{Li}},
  \bibinfo{journal}{Phys. Rev. C} \textbf{\bibinfo{volume}{72}},
  \bibinfo{pages}{064309} (\bibinfo{year}{2005}).

\bibitem[{\citenamefont{Agrawal et~al.}(2005)\citenamefont{Agrawal, Shlomo, and
  Au}}]{Agrawal05}
\bibinfo{author}{\bibfnamefont{B.~K.} \bibnamefont{Agrawal}},
  \bibinfo{author}{\bibfnamefont{S.}~\bibnamefont{Shlomo}}, \bibnamefont{and}
  \bibinfo{author}{\bibfnamefont{V.~K.} \bibnamefont{Au}},
  \bibinfo{journal}{Phys. Rev. C} \textbf{\bibinfo{volume}{72}},
  \bibinfo{pages}{014310} (\bibinfo{year}{2005}).

\bibitem[{\citenamefont{Kumar et~al.}(2006)\citenamefont{Kumar, Agrawal, and
  Dhiman}}]{Kumar06}
\bibinfo{author}{\bibfnamefont{R.}~\bibnamefont{Kumar}},
  \bibinfo{author}{\bibfnamefont{B.~K.} \bibnamefont{Agrawal}},
  \bibnamefont{and} \bibinfo{author}{\bibfnamefont{S.~K.}
  \bibnamefont{Dhiman}}, \bibinfo{journal}{Phys. Rev. C}
  \textbf{\bibinfo{volume}{74}}, \bibinfo{pages}{034323}
  (\bibinfo{year}{2006}).

\bibitem[{\citenamefont{Bednarek and Manka}(2005)}]{Bednarek05}
\bibinfo{author}{\bibfnamefont{I.}~\bibnamefont{Bednarek}} \bibnamefont{and}
  \bibinfo{author}{\bibfnamefont{R.}~\bibnamefont{Manka}}, \bibinfo{journal}{J.
  Phys. G} \textbf{\bibinfo{volume}{31}}, \bibinfo{pages}{1009}
  (\bibinfo{year}{2005}).

\bibitem[{\citenamefont{Schaffner and Gal}(2000)}]{Schaffner00}
\bibinfo{author}{\bibfnamefont{J.}~\bibnamefont{Schaffner}} \bibnamefont{and}
  \bibinfo{author}{\bibfnamefont{A.}~\bibnamefont{Gal}},
  \bibinfo{journal}{Phys. Rev. C} \textbf{\bibinfo{volume}{62}},
  \bibinfo{pages}{034311} (\bibinfo{year}{2000}).

\bibitem[{\citenamefont{Glendenning and Moszkowski}(1991)}]{Glendenning91}
\bibinfo{author}{\bibfnamefont{N.~K.} \bibnamefont{Glendenning}}
  \bibnamefont{and} \bibinfo{author}{\bibfnamefont{S.~A.}
  \bibnamefont{Moszkowski}}, \bibinfo{journal}{Phys. Rev. Lett.}
  \textbf{\bibinfo{volume}{67}}, \bibinfo{pages}{2414} (\bibinfo{year}{1991}).

\bibitem[{\citenamefont{Lalazissis et~al.}(1997)\citenamefont{Lalazissis,
  Konig, and Ring}}]{Lalazissis97}
\bibinfo{author}{\bibfnamefont{G.~A.} \bibnamefont{Lalazissis}},
  \bibinfo{author}{\bibfnamefont{J.}~\bibnamefont{Konig}}, \bibnamefont{and}
  \bibinfo{author}{\bibfnamefont{P.}~\bibnamefont{Ring}},
  \bibinfo{journal}{Phys. Rev. C} \textbf{\bibinfo{volume}{55}},
  \bibinfo{pages}{540} (\bibinfo{year}{1997}).

\bibitem[{\citenamefont{Weinberg}(1972)}]{Weinberg72}
\bibinfo{author}{\bibfnamefont{S.}~\bibnamefont{Weinberg}},
  \emph{\bibinfo{title}{Gravitation and Cosmology}} (\bibinfo{publisher}{Wiley,
  New York}, \bibinfo{year}{1972}).

\bibitem[{\citenamefont{Baym et~al.}(1971)\citenamefont{Baym, Pethick, and
  Sutherland}}]{Baym71}
\bibinfo{author}{\bibfnamefont{G.}~\bibnamefont{Baym}},
  \bibinfo{author}{\bibfnamefont{C.}~\bibnamefont{Pethick}}, \bibnamefont{and}
  \bibinfo{author}{\bibfnamefont{P.}~\bibnamefont{Sutherland}},
  \bibinfo{journal}{Astrophys. J.} \textbf{\bibinfo{volume}{170}},
  \bibinfo{pages}{299} (\bibinfo{year}{1971}).

\bibitem[{\citenamefont{Lattimer et~al.}(1990)\citenamefont{Lattimer, Prakash,
  Masak, and Yahil}}]{Lattimer90}
\bibinfo{author}{\bibfnamefont{J.~M.} \bibnamefont{Lattimer}},
  \bibinfo{author}{\bibfnamefont{M.}~\bibnamefont{Prakash}},
  \bibinfo{author}{\bibfnamefont{D.}~\bibnamefont{Masak}}, \bibnamefont{and}
  \bibinfo{author}{\bibfnamefont{A.}~\bibnamefont{Yahil}},
  \bibinfo{journal}{Astrophys. J.} \textbf{\bibinfo{volume}{355}},
  \bibinfo{pages}{241} (\bibinfo{year}{1990}).

\bibitem[{\citenamefont{Lattimer and Prakash}(2001)}]{Lattimer01}
\bibinfo{author}{\bibfnamefont{J.~M.} \bibnamefont{Lattimer}} \bibnamefont{and}
  \bibinfo{author}{\bibfnamefont{M.}~\bibnamefont{Prakash}},
  \bibinfo{journal}{Astrophys. J.} \textbf{\bibinfo{volume}{550}},
  \bibinfo{pages}{426} (\bibinfo{year}{2001}).

\bibitem[{\citenamefont{Lattimer and Prakash}(2004)}]{Lattimer04}
\bibinfo{author}{\bibfnamefont{J.~M.} \bibnamefont{Lattimer}} \bibnamefont{and}
  \bibinfo{author}{\bibfnamefont{M.}~\bibnamefont{Prakash}},
  \bibinfo{journal}{Science} \textbf{\bibinfo{volume}{304}},
  \bibinfo{pages}{536} (\bibinfo{year}{2004}).

\bibitem[{\citenamefont{Schaffner and Mishustin}(1996)}]{Schaffner96}
\bibinfo{author}{\bibfnamefont{J.}~\bibnamefont{Schaffner}} \bibnamefont{and}
  \bibinfo{author}{\bibfnamefont{I.~N.} \bibnamefont{Mishustin}},
  \bibinfo{journal}{Phys. Rev. C} \textbf{\bibinfo{volume}{53}},
  \bibinfo{pages}{1416} (\bibinfo{year}{1996}).

\bibitem[{\citenamefont{Shen}(2002)}]{Shen02}
\bibinfo{author}{\bibfnamefont{H.}~\bibnamefont{Shen}}, \bibinfo{journal}{Phys.
  Rev. C} \textbf{\bibinfo{volume}{65}}, \bibinfo{pages}{035802}
  (\bibinfo{year}{2002}).

\bibitem[{\citenamefont{Bejger et~al.}(2005)\citenamefont{Bejger, Bulik, and
  Haensel}}]{Bejger05}
\bibinfo{author}{\bibfnamefont{M.}~\bibnamefont{Bejger}},
  \bibinfo{author}{\bibfnamefont{T.}~\bibnamefont{Bulik}}, \bibnamefont{and}
  \bibinfo{author}{\bibfnamefont{P.}~\bibnamefont{Haensel}},
  \bibinfo{journal}{Mon. Not. R. Astron. Soc.} \textbf{\bibinfo{volume}{364}},
  \bibinfo{pages}{635} (\bibinfo{year}{2005}).

\bibitem[{\citenamefont{Heinke et~al.}(2006)\citenamefont{Heinke, Rybicki,
  Narayan, and Grindlay}}]{Heinke06}
\bibinfo{author}{\bibfnamefont{C.~O.} \bibnamefont{Heinke}},
  \bibinfo{author}{\bibfnamefont{G.~B.} \bibnamefont{Rybicki}},
  \bibinfo{author}{\bibfnamefont{R.}~\bibnamefont{Narayan}}, \bibnamefont{and}
  \bibinfo{author}{\bibfnamefont{J.~E.} \bibnamefont{Grindlay}},
  \bibinfo{journal}{Astrophys. J.} \textbf{\bibinfo{volume}{644}},
  \bibinfo{pages}{1090} (\bibinfo{year}{2006}).

\bibitem[{\citenamefont{Ozel}(2006)}]{Ozel06}
\bibinfo{author}{\bibfnamefont{F.}~\bibnamefont{Ozel}},
  \bibinfo{journal}{Nature} \textbf{\bibinfo{volume}{441}},
  \bibinfo{pages}{1115} (\bibinfo{year}{2006}).

\bibitem[{\citenamefont{Villarreal and Strohmayer}(2004)}]{Villarreal04}
\bibinfo{author}{\bibfnamefont{A.~R.} \bibnamefont{Villarreal}}
  \bibnamefont{and} \bibinfo{author}{\bibfnamefont{T.~E.}
  \bibnamefont{Strohmayer}}, \bibinfo{journal}{Astrophys. J.}
  \textbf{\bibinfo{volume}{614}}, \bibinfo{pages}{L121} (\bibinfo{year}{2004}).

\bibitem[{\citenamefont{Y.Sugahara and
  H.Toki}(1994{\natexlab{b}})}]{Sugahara94a}
\bibinfo{author}{\bibnamefont{Y.Sugahara}} \bibnamefont{and}
  \bibinfo{author}{\bibnamefont{H.Toki}}, \bibinfo{journal}{Prog. Theor. Phys}
  \textbf{\bibinfo{volume}{92}}, \bibinfo{pages}{803}
  (\bibinfo{year}{1994}{\natexlab{b}}).


\bibitem[{\citenamefont{Greco et~al.}(2001)\citenamefont{Greco, Matera,
  Colonna, Toro, and Fabbri}}]{Greco01}
\bibinfo{author}{\bibfnamefont{V.}~\bibnamefont{Greco}},
  \bibinfo{author}{\bibfnamefont{F.}~\bibnamefont{Matera}},
  \bibinfo{author}{\bibfnamefont{M.}~\bibnamefont{Colonna}},
  \bibinfo{author}{\bibfnamefont{M.~Di} \bibnamefont{Toro}}, \bibnamefont{and}
  \bibinfo{author}{\bibfnamefont{G.}~\bibnamefont{Fabbri}},
  \bibinfo{journal}{Phys. Rev.C} \textbf{\bibinfo{volume}{63}},
  \bibinfo{pages}{035202} (\bibinfo{year}{2001}).

\bibitem[{\citenamefont{Panda et~al.}(2006)\citenamefont{Panda, da~Providência,
  and Providência}}]{Panda06}
\bibinfo{author}{\bibfnamefont{P.~K.} \bibnamefont{Panda}},
  \bibinfo{author}{\bibfnamefont{J.}~\bibnamefont{da~Providência}},
  \bibnamefont{and}
  \bibinfo{author}{\bibfnamefont{C.}~\bibnamefont{Providência}},
  \bibinfo{journal}{Phys. Rev. C} \textbf{\bibinfo{volume}{73}},
  \bibinfo{pages}{035805} (\bibinfo{year}{2006}).

\bibitem[{\citenamefont{Shapiro and Teukolsky}(1983)}]{Shapiro83}
\bibinfo{author}{\bibfnamefont{S.~L.} \bibnamefont{Shapiro}} \bibnamefont{and}
  \bibinfo{author}{\bibfnamefont{S.~A.} \bibnamefont{Teukolsky}},
  \emph{\bibinfo{title}{Black Holes, White Dwarfs, and Neutron Stars}}
  (\bibinfo{publisher}{Wiley, New York}, \bibinfo{year}{1983}).

\bibitem[{\citenamefont{Stergioulas and Friedman}(1995)}]{Stergioulas95}
\bibinfo{author}{\bibfnamefont{N.}~\bibnamefont{Stergioulas}} \bibnamefont{and}
  \bibinfo{author}{\bibfnamefont{J.~L.} \bibnamefont{Friedman}},
  \bibinfo{journal}{Astrophys. J.} \textbf{\bibinfo{volume}{444}},
  \bibinfo{pages}{306} (\bibinfo{year}{1995}).

\bibitem[{\citenamefont{Bejger et~al.}(2007)\citenamefont{Bejger, Haensel, and
  Zdunik}}]{Bejger06}
\bibinfo{author}{\bibfnamefont{M.}~\bibnamefont{Bejger}},
  \bibinfo{author}{\bibfnamefont{P.}~\bibnamefont{Haensel}}, \bibnamefont{and}
  \bibinfo{author}{\bibfnamefont{J.}~\bibnamefont{Zdunik}},
  \bibinfo{journal}{Astro.Astrophys.} \textbf{\bibinfo{volume}{464}},
  \bibinfo{pages}{L49} (\bibinfo{year}{2007}).

\bibitem[{\citenamefont{Danielewicz et~al.}(2002)\citenamefont{Danielewicz,
  Lacey, and Lynch}}]{Danielewicz02}
\bibinfo{author}{\bibfnamefont{P.}~\bibnamefont{Danielewicz}},
  \bibinfo{author}{\bibfnamefont{R.}~\bibnamefont{Lacey}}, \bibnamefont{and}
  \bibinfo{author}{\bibfnamefont{G.}~\bibnamefont{Lynch}},
  \bibinfo{journal}{Science} \textbf{\bibinfo{volume}{298}},
  \bibinfo{pages}{1592} (\bibinfo{year}{2002}).

\end{thebibliography}

\newpage
\begin{figure}
\caption{\label{fig:xsig_xome} (Color online) Variations of $X_{\sigma
Y}$ with $X_{\omega Y}$ for $\Lambda$, $\Sigma$ and $\Xi$  hyperons.
The values of $X_{\sigma Y}$ for a given value of $X_{\omega Y}$
are calculated using Eqs.  (\ref{eq:uyn} and \ref{eq:xw}).  }
\caption{\label{fig:jl_skin} (Color online) Variations of the symmetry
energy coefficient $J$ (upper panel) and its linear density dependence $L$
(lower panel) as a function of $\Delta r$ for different parameterizations
with $\zeta = 0.0, 0.03$ and $0.06$ . }

\caption{\label{fig:error_br} (Color online) Variations of the rms errors
in the total binding energies (upper panel) and charge rms radii (lower panel)
as a function of $\Delta r$ for different parameterizations with $\zeta
=0.0, 0.03$ and $0.06$.  }

\caption{\label{fig:eos}  (Color online) The EOSs for pure neutron
matter (upper panel) and symmetric nuclear matter (lower panel). The
solid and dashed curves correspond  to $\Delta r$ = 0.16 and 0.28 fm,
respectively. The shaded regions represent the experimental data taken
from Ref. \cite{Danielewicz02}.}

\caption{\label{fig:mu} (Color online) The chemical potentials for neutron
(upper panel) and electron (lower panel) as a function of density.}

\caption{\label{fig:m-r} (Color online) Variation of  neutron star mass as
a function of its radius $R$ for selected EOSs. These EOSs are obtained
using  the parameter sets LY (UY) correspond  to combinations of  $\Delta r$
= 0.16(0.28), $\zeta = 0.06(0.0)$ and $X_{\omega Y} = 0.5 (0.8)$. The
parameter sets L0 and U0 are analogous to LY and UY respectively, but,
with no hyperons.  The various constraints as indicated by causality,
rotation, and $\Delta I/I = 0.014$ are discussed in the text.}

\caption{\label{fig:thrd} (Color online) The threshold density for various
hyperons and central densities for neutron star with the canonical mass
and maximum mass obtained for the EOSs corresponding to the  parameter
sets LY and UY.}

\caption{\label{fig:frac}(Color online)
Particle  fractions as a function of radial coordinate of the neutron
star obtained at maximum mass for LY (upper panel) and UY (lower panel)
parameterizations. The curves labeled as "n/3" should be multiplied by
three to get the  actual neutron fractions.}

\end{figure}

\begin{figure}
\caption{\label{fig:r14_z14} (Color online) Variations of radius
($R_{1.4}$) and redshift ($Z_{1.4}$) for  neutron star with the canonical
mass as a function of maximum neutron star mass obtained for the EOSs
corresponding to the all different parameterizations of the extended FTRMF
model as considered.  The vertical line at $M_{\text{max}} =1.6M_\odot$
in the upper panel corresponds to the mass of the PSR J0751+1807 measured
with $95\%$ confidence limit.}

\caption{\label{fig:rmax_zmax} (Color online)
Variations of radius ($R_{\text{max}}$) and redshift ($Z_{\text
{max}}$) for  neutron star with the maximum  mass as a function of
maximum neutron star mass obtained for the EOSs corresponding to the all
different parameterizations of the extended FTRMF model as considered.
The horizontal line in the lower panel corresponds to the measured
value of the redshift, $Z = 0.35$, for the neutron star EXO 0748-676
\cite{Ozel06}.}

\caption{\label{fig:kep_seq} (Color online) 
The relationship between mass M and the circumferential equatorial radius
$R_{\text{eq}}$  for Keplerian sequences for different EOSs obtained
within the extended FTRMF model.}

\caption{\label{fig:mreq_1122} (Color online)
The mass M  verses the circumferential equatorial radius  $R_{\text{eq}}$
for the neutron stars rotating at $1122$ Hz for selected EOSs obtained
within the extended FTRMF model. The minimum values for the radius indicated
by open circles are determined by the setting-in of the instability
with respect to axi-symmetric perturbations. The maximum values for the
radius  indicated by open squares are determined by the mass-shedding
instability.  The values of maximum radius are well fitted by solid
curve obtained using Eq.(\ref{eq:rmax}).}

\end{figure}
\newpage

\begin{sidewaystable}
\caption{\label{tab:para1}
New coupling strength parameters for the Lagrangian of the extended
FTRMF model as given in Eq.(\ref{eq:lden}).  The seven different
parameter sets  correspond to the different values of the neutron
skin-thickness $\Delta r$ for the $^{208}$Pb nucleus used in the
fit. The value of $\omega$-meson self-coupling $\zeta $ is equal to
0.0 for all these parameterizations. The values of $\Delta r$ are
in fm, the parameters $\overline{\kappa}$, $\overline{\alpha_1}$,
and $\overline{\alpha{_2}}$ are in  fm$^{-1}$ and $m_{\sigma}$ are in
MeV. The masses for other mesons are taken to be $m_{\omega}=782.5$ MeV,
$m_{\rho}=763$ MeV, $m_{\sigma}^*=975$ MeV and $m_\phi=1020$  MeV.
For the masses of  nucleons and hyperons  we use $M_N=939$ MeV, $M_\Lambda
=1116$ MeV, $M_\Sigma =1193$ MeV and $M_\Xi =1313$ MeV.  The values of
$\overline{\kappa}$, $\overline{\lambda}$, $\overline{\alpha_1}$,
$\overline{\alpha_1}^{\prime}$, $\overline{\alpha_2}$, 
$\overline{\alpha}_2^{\prime}$, and $\overline{\alpha}_3^{\prime}$
are multiplied with $10^{2}$.}

\begin{ruledtabular}
\begin{tabular}{|cddddddd|}
\multicolumn{1}{|c}{${\Delta r}$}&
\multicolumn{1}{c}{ 0.16}&
\multicolumn{1}{c}{ 0.18}&
\multicolumn{1}{c}{ 0.20}&
\multicolumn{1}{c}{ 0.22}&
\multicolumn{1}{c}{ 0.24}&
\multicolumn{1}{c}{ 0.26}&
\multicolumn{1}{c|}{ 0.28}\\
\hline
$ g_{\sigma N}$ &10.51369 &10.65616 & 10.44426&10.50339 &10.34061 &10.48597 &10.32009\\
$ g_{\omega N}$ &13.48789 &13.95799 &13.52239 & 13.80084&13.46209&13.81202 &13.45113\\
$ g_{\rho N }$ &14.98497 &14.32687 &13.11709 &12.12975 &11.18278 &10.39449 &10.09608\\
$ \overline {\kappa}$&2.62556 &3.02154 &2.43049 &3.39711 &3.24752 &3.05611 &2.82791\\
$ \overline {\lambda}$&-0.73495&-0.45437 &-0.04279 &-1.15784&-1.36867 & -0.86772&-1.13890\\
$ \overline{\alpha}_1$& 0.22672&0.38665 &0.18121 &0.44021 &0.35304 &0.34843 &0.23357\\
$ \overline{\alpha}_1^{\prime}$ &0.07325 &0.07791 &0.15979 &0.00987 &0.00725 &0.052231 &0.04733\\
$ \overline{\alpha}_2$ &3.05925 &2.91796 &2.96668 &2.56759 &2.27472 &0.68086 &0.60739\\
$ \overline{\alpha}_2^{\prime}$ &1.55587 &1.35016 &1.25303 &0.51396 &0.15515 &0.41389 &0.33057 \\
$ \overline{\alpha}_3^{\prime}$ &1.50060 &1.47585 &0.09727 &1.04562 &0.52777 &1.14566 &0.30434\\
$ m_{\sigma}$&502.23217 &495.76339 &497.83489 &491.48257 &492.76821 &490.24238 &491.86681 \\
\hline
\end{tabular}
\end{ruledtabular}
\end{sidewaystable}

\begin{sidewaystable}
\caption{\label{tab:para2}
Same as Table \ref{tab:para1}, but, with $\omega$-meson self coupling $\zeta = 0.03$.} 

\begin{ruledtabular}
\begin{tabular}{|cddddddd|}
\multicolumn{1}{|c}{$ {\Delta r}$}&
\multicolumn{1}{c}{ 0.16}&
\multicolumn{1}{c}{ 0.18}&
\multicolumn{1}{c}{ 0.20}&
\multicolumn{1}{c}{ 0.22}&
\multicolumn{1}{c}{ 0.24}&
\multicolumn{1}{c}{ 0.26}&
\multicolumn{1}{c|}{ 0.28}\\
\hline
$ g_{\sigma N}$ &10.62886 &10.76147 &10.73005 &10.71942 &10.61808 &10.67656 &10.60110\\
$ g_{\omega N}$ &13.65991 &14.11102&14.04275 &14.12534 &13.88708 &14.11958 &14.03101\\
$ g_{\rho N }$ &14.99076 &14.67414 &13.69014 &12.19156 &10.96456 & 10.14811&10.00441\\
$ \overline {\kappa}$ &1.38118 &1.56065 &1.62316 &1.61820 &1.77184 & 1.68916&1.78793\\
$ \overline {\lambda}$ &0.58536 &0.97528 &0.64498 &1.06102 &0.48269 &0.86649 &0.74676\\
$ \overline{\alpha}_{1}$&0.00366 &0.10311 &0.08281 &0.10650 &0.12586 & 0.11999&0.16088\\
$  \overline{\alpha}_1^{\prime}$ &0.02717 &0.05071 &0.02980 &0.06526 &0.00052 &0.04411 &0.01669\\
$ \overline{\alpha}_{2}$ &2.89393 &3.06821 &3.18222 &2.77747 &1.18745 &0.68168 &0.47146\\
$ \overline{\alpha}_2^{\prime}$ &1.59659 &1.16255 &0.47540 &0.22126 &1.27574 &0.54787 &0.52816 \\
$ \overline{\alpha}_3^{\prime}$ &1.52088 &1.35981 &0.97721 &0.45581 &0.28975 & 0.35906&0.32358\\
$ m_{\sigma}$&506.50582 &500.51106 &499.52635 & 497.20745&499.12460 & 495.18211&494.93882 \\
\hline
\end{tabular}
\end{ruledtabular}
\end{sidewaystable}

\begin{sidewaystable}
\caption{\label{tab:para3} 
Same as Table \ref{tab:para1}, but, with $\omega$-meson self coupling $\zeta =0.06$.} 
\begin{ruledtabular}
\begin{tabular}{|cddddddd|}
\multicolumn{1}{|c}{$ {\Delta r}$}&
\multicolumn{1}{c}{ 0.16}&
\multicolumn{1}{c}{ 0.18}&
\multicolumn{1}{c}{ 0.20}&
\multicolumn{1}{c}{ 0.22}&
\multicolumn{1}{c}{ 0.24}&
\multicolumn{1}{c}{ 0.26}&
\multicolumn{1}{c|}{ 0.28}\\
\hline
$ g_{\sigma N}$ &11.05170 &11.02412 &10.95765 &11.01908 &10.91944 &11.10806 &11.03151\\
$ g_{\omega N}$ &14.65579 &14.66595 &14.59582 &14.77458 &14.64700 &15.19792 &15.01572\\
$ g_{\rho N }$ &14.98725&14.52186 &13.41111 &11.94837&10.71055 & 10.08835&10.00666\\
$ \overline {\kappa}$ &0.66576 &0.69497 &0.76852 &0.78002 &0.90221 &1.13349 &0.80797\\
$ \overline {\lambda}$ &2.46427 &2.44874 &2.41259 &2.47238 &2.33265 &2.51229 &2.41320\\
$ \overline{\alpha}_{1}$&0.00601  & 0.00449&0.00409  &0.01469 &0.03499 &0.14153 &0.02073\\
$  \overline{\alpha}_1^{\prime}$ &0.00203 &0.00526 &0.01079 &0.01559 &0.00230 &0.00085 &0.01109\\
$ \overline{\alpha}_{2}$ &2.86236 &2.58355 &2.66308 &2.02292 &1.24695 &1.18538  &0.55325\\
$ \overline{\alpha}_2^{\prime}$ &1.55176&1.56881&1.30876&0.90169 &0.77919&0.27422 &0.16326\\
$ \overline{\alpha}_3^{\prime}$ &1.55307 &1.58487 &0.84916 &0.96305 & 0.74863&0.40699 &0.72768\\
$ m_{\sigma}$&503.43838 &501.37038 &499.38134&497.27203&495.82388 &490.83495 & 490.68907\\
\hline
\end{tabular}
\end{ruledtabular}
\end{sidewaystable}

\begin{table}[p]
\caption{
The values of central baryon density $\rho_c$, mass $M$, radius $R$ ,
radiation radius $R_\infty$, binding energy $E_{\text{bind}}$ and redshift
$Z$ for non-rotating neutron stars with maximum mass calculated for the
EOSs obtained using LY, UY, L0 and U0 parameterizations.  The parameter
sets LY (L0) and UY (U0) yield softest and stiffest EOS with (without)
hyperons in comparison to all other parameterizations obtained in
Sec. \ref{sec:paramet}.}

\begin{ruledtabular}
\begin{tabular}{|cdddd|}
&  LY & UY & L0 & U0\\
\hline
$\rho_c$ (fm$^{-3}$)&  1.05 & 0.84 & 1.12&0.79\\
$M (M_{\odot})$&  1.4 &2.1  &1.7 &2.4\\
$R$(km)&11.3   &12.0  &10.9& 12.2\\
$R_{\infty}$(km)&14.2   &17.3  &14.9  &18.9\\
$E_{{\text bind}}$ (10$^{53}$ergs)  & 1.36 & 3.80 &2.76&6.49\\
$Z$& 0.25 &0.41  &0.37  &0.57\\
\end{tabular}
\end{ruledtabular}
\end{table}

\begin{table}[p]
\caption{ Same as Table IV, but, for the non-rotating neutron stars with canonical mass.}

\begin{ruledtabular}
\begin{tabular}{|cdddd|}
&  LY & UY & L0 & U0\\
\hline
$\rho_{c} $(fm$^{-3}$)&1.05  &0.32& 0.50  &0.32\\
 $R $(km)& 11.3& 14.1  & 12.5 &14.1\\
$R_{\infty}$(km)&14.2   &16.8  &15.3  &16.8\\
$E_{{\text bind}} $($10^{53}$ergs)&1.36&1.10  & 1.37  &1.10\\
$Z$& 0.25  & 0.19&0.22  &0.19\\
\end{tabular}
\end{ruledtabular}
\end{table}

\newpage
\begin{table}[p]
\caption{ Values of the $M_{\text{max}}$, $R_{\text {1.4}}$ and the
radius $R_{\text{max}}$ for the neutron star with maximum  mass obtained
for the EOSs corresponding to the  selected combinations of $\Delta r$,
$\zeta$ and $X_{\omega Y}$.}

\begin{ruledtabular}
\begin{tabular}{|cccccccc|}
\hline
\multicolumn{1}{|c}{}&
\multicolumn{1}{c}{}&
\multicolumn{3}{c}{$\zeta=0.0$}&
\multicolumn{3}{c|}{$\zeta=0.06$}\\
\cline{3-8}
$X_{\omega Y}$& $\Delta r$& $M_{\text{max}}$ & $R_{1.4}$& $R_{\text{max}}$
& $M_{\text{max}}$& $R_{1.4}$& $R_{\text{max}}$\\
&(fm)& ($M_{\odot}$)& (km) &(km)&($M_{\odot}$)& (km) &(km)\\
\hline
0.50& 0.16& 1.8& 13.4& 12.0& 1.4& 11.3& 11.3\\
  & 0.28& 1.8& 14.1& 12.2& 1.4& 11.6& 11.6\\
\hline
0.80& 0.16& 2.1& 13.4& 12.0& 1.5& 12.3& 11.0\\
 & 0.28& 2.1& 14.1& 12.1& 1.5& 13.0& 11.3\\
\end{tabular}
\end{ruledtabular}
\end{table}

\newpage
\begin{table}[p]
\caption{\label{tab:ns_1122}The properties of neutron star rotating with
1122 Hz for different EOSs calculated within FTRMF model.}

\begin{ruledtabular}
\begin{tabular}{|ccccc|}
EOS&$M(R_{\text {eq}}^{{\text min}}$)&$R_{\text {eq}}^{{\text min}}$&r$_{{\text
pole}}$/r$_{{\text eq}}$&$T/|W|$\\ 
&($M\odot$)&(km)& &\\
\hline
L0&1.908  &12.12 & 0.804 &0.054 \\
U0& 2.721  &13.46 & 0.815 & 0.063\\
LY& 1.624 & 13.85& 0.692 &0.076 \\
UY& 2.266  &12.99 & 0.799 & 0.059\\
\hline
EOS&$M(R_{\text {eq}}^{{\text{max}}}$)&$R_{\text {eq}}^{{\text{max}}}$&r$_{{\text
{pole}}}$/r$_{{\text {eq}}}$&$T/| W|$\\
&($M\odot$)&(km)& &\\ 
\hline
L0&1.909  &17.14 & 0.566 &0.109 \\
U0& 2.556  &18.73 & 0.556 & 0.127\\
LY& 1.694 & 16.49& 0.575 &0.097 \\
UY& 2.360  &18.35 & 0.559 & 0.118\\
\end{tabular}
\end{ruledtabular}
\end{table}

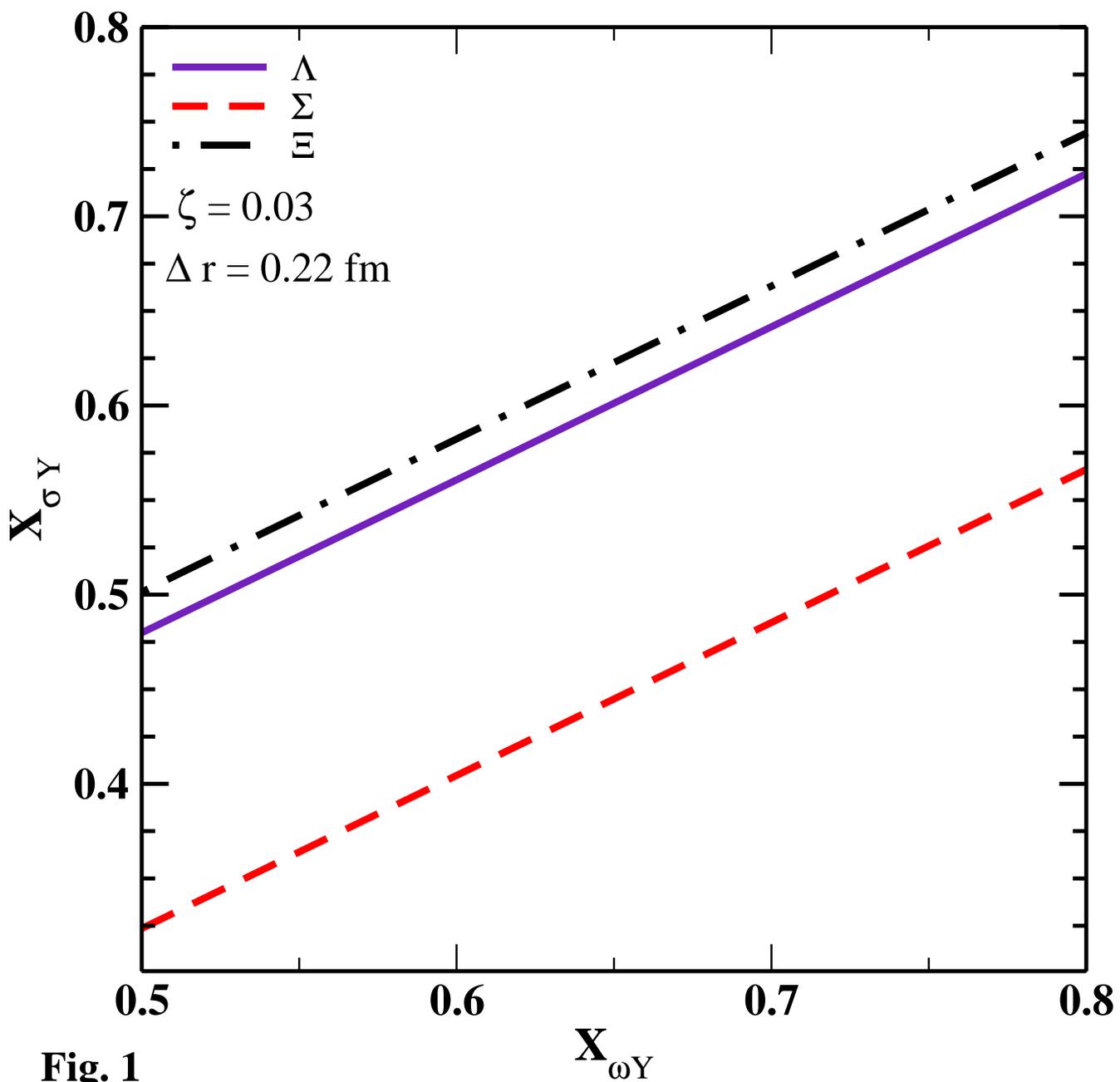
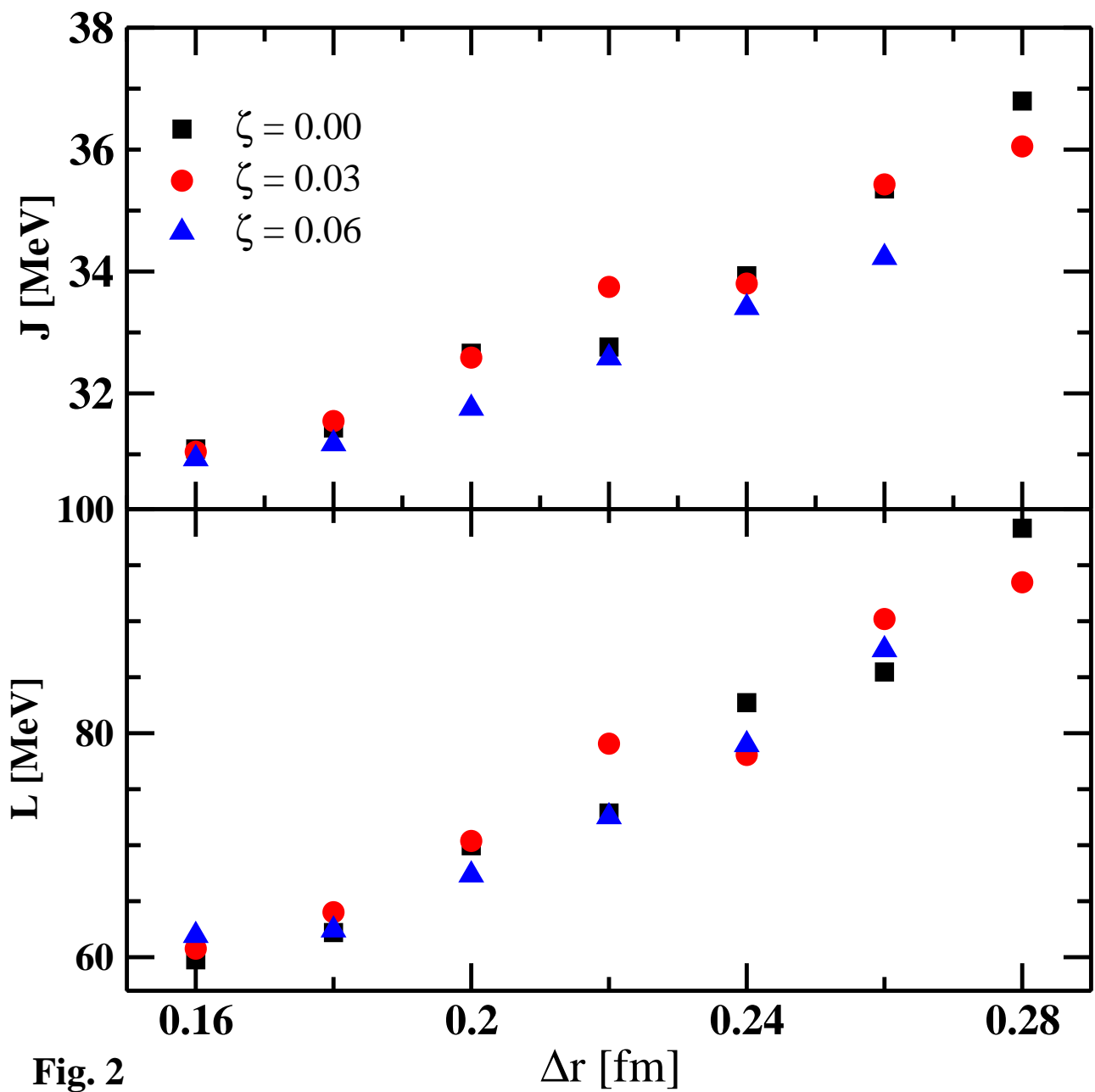
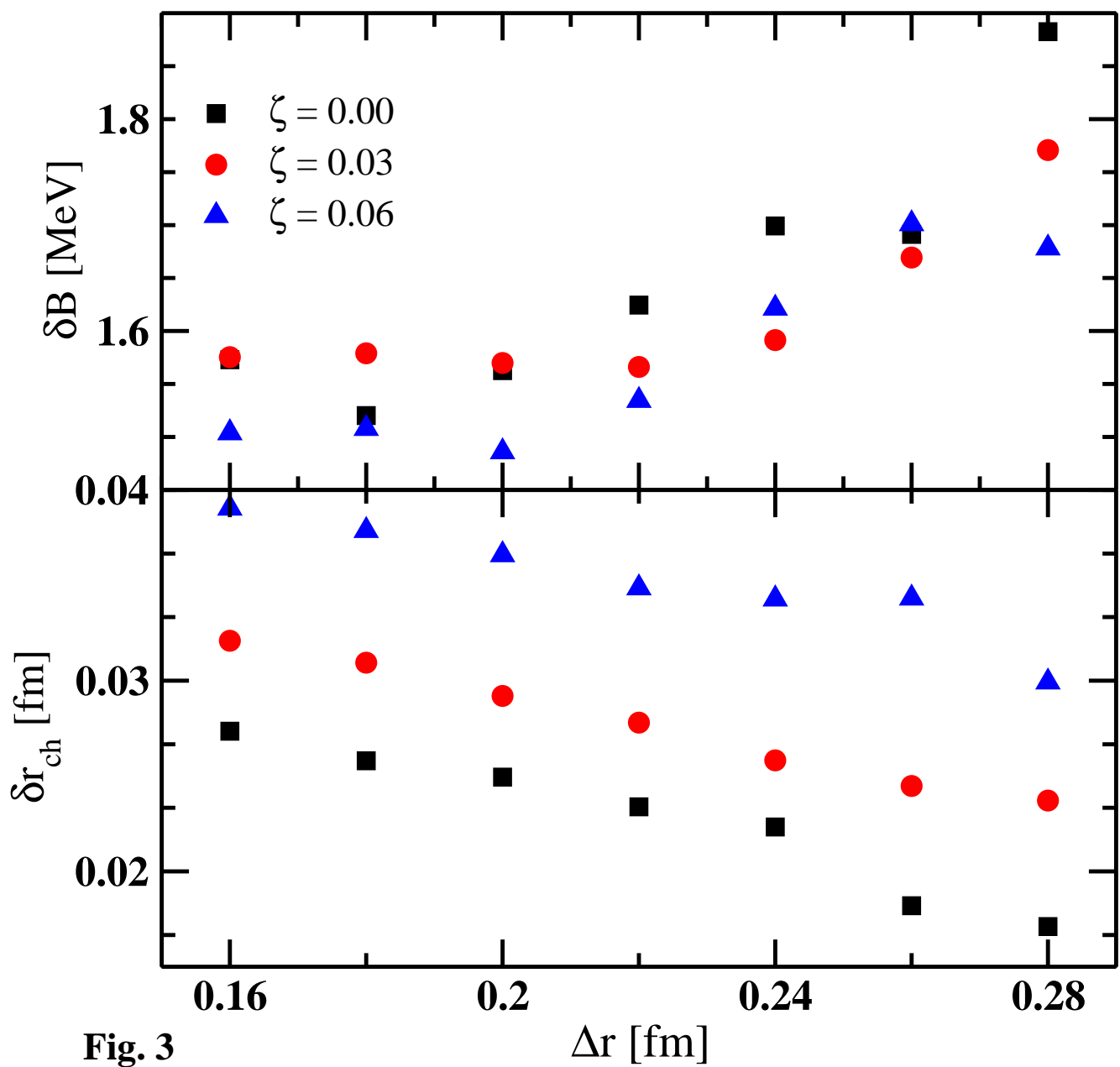
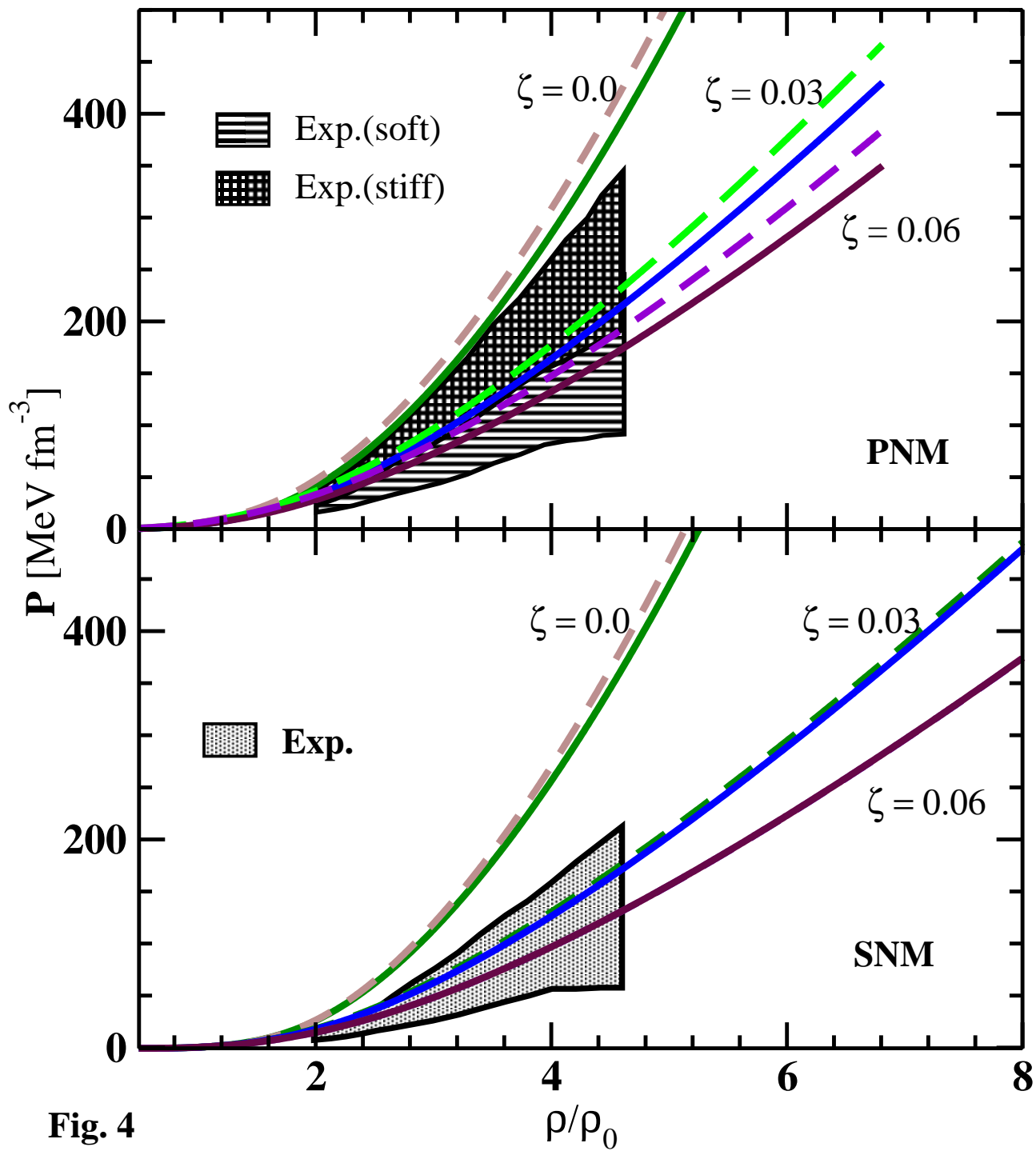
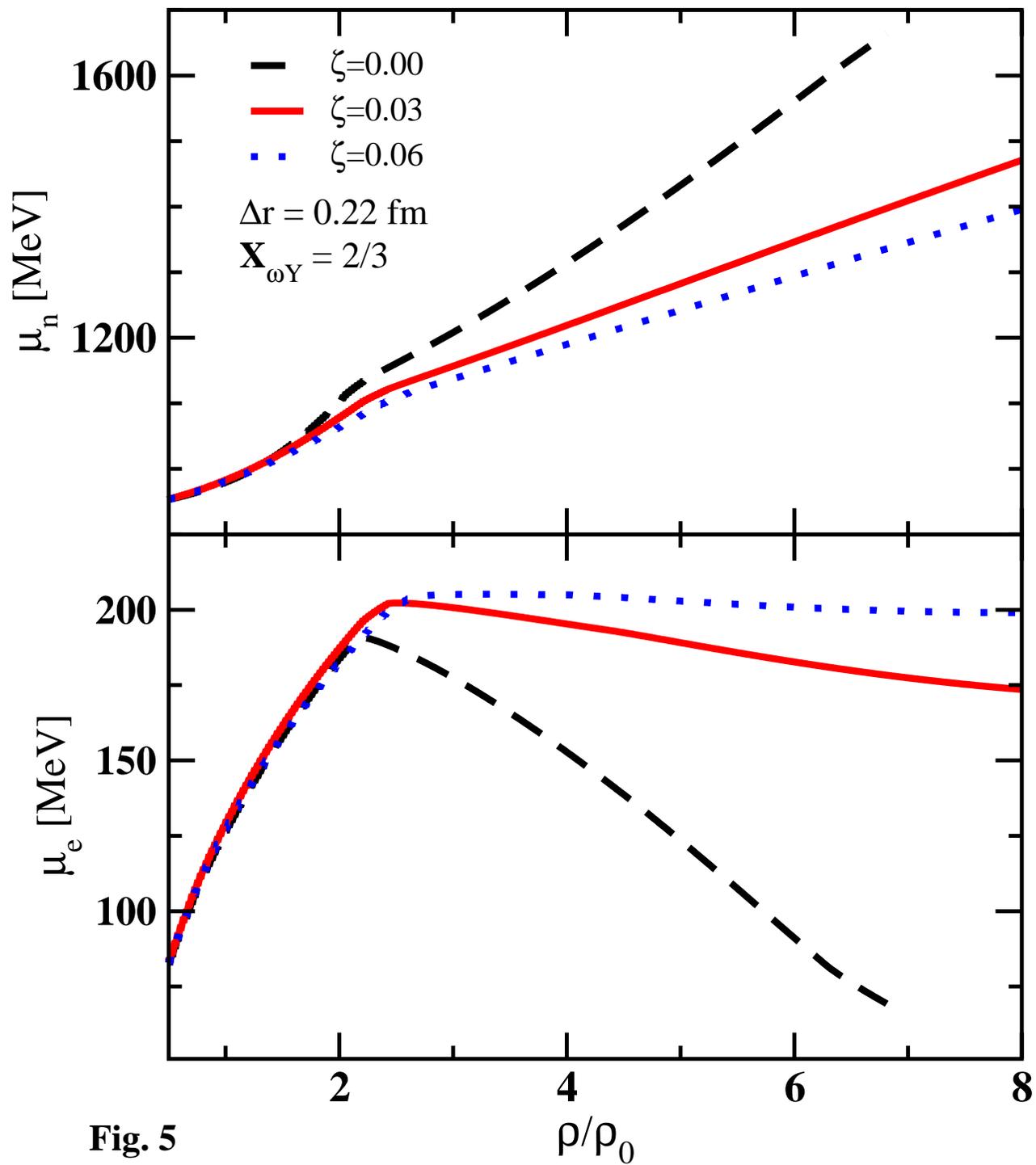
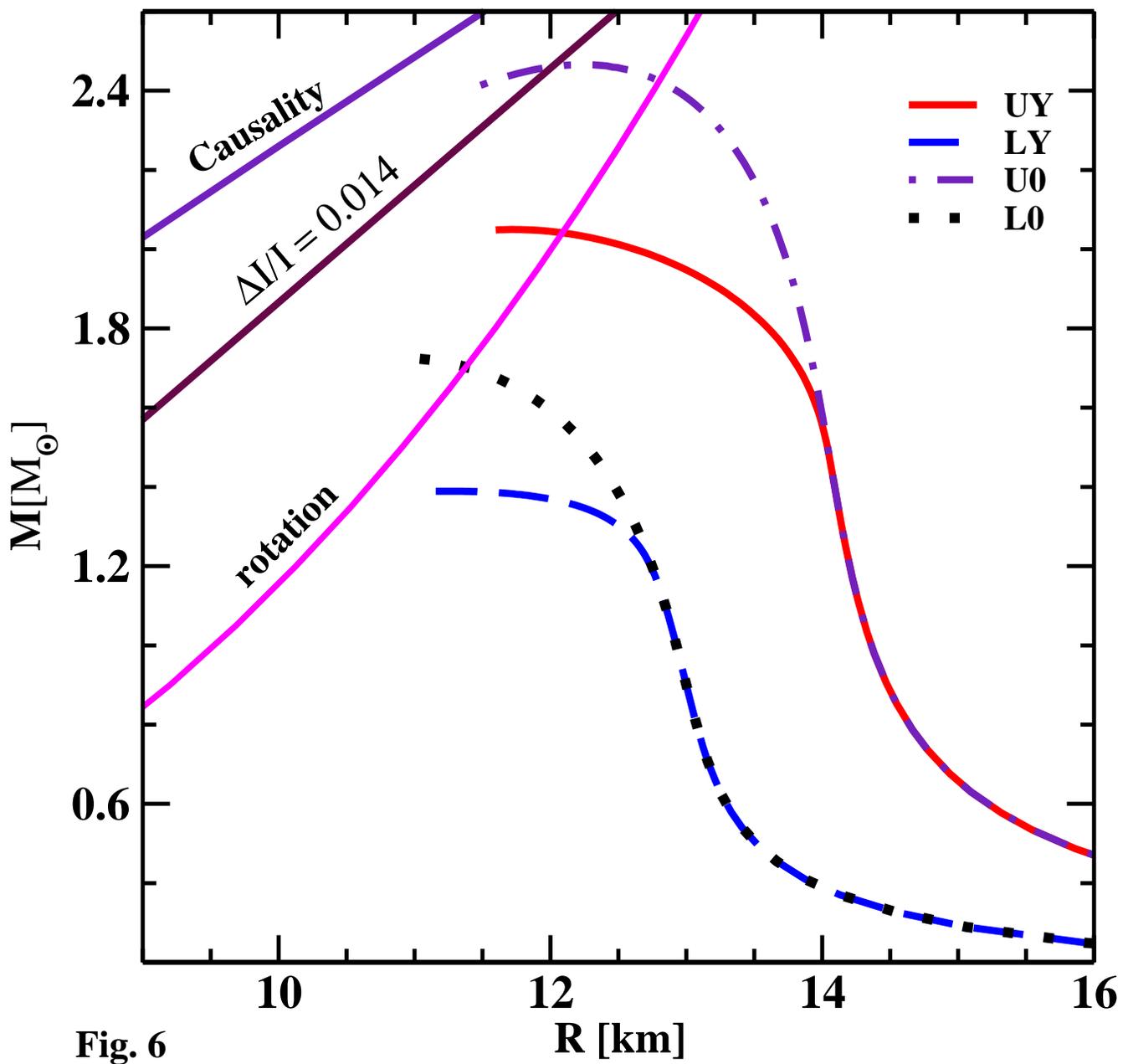
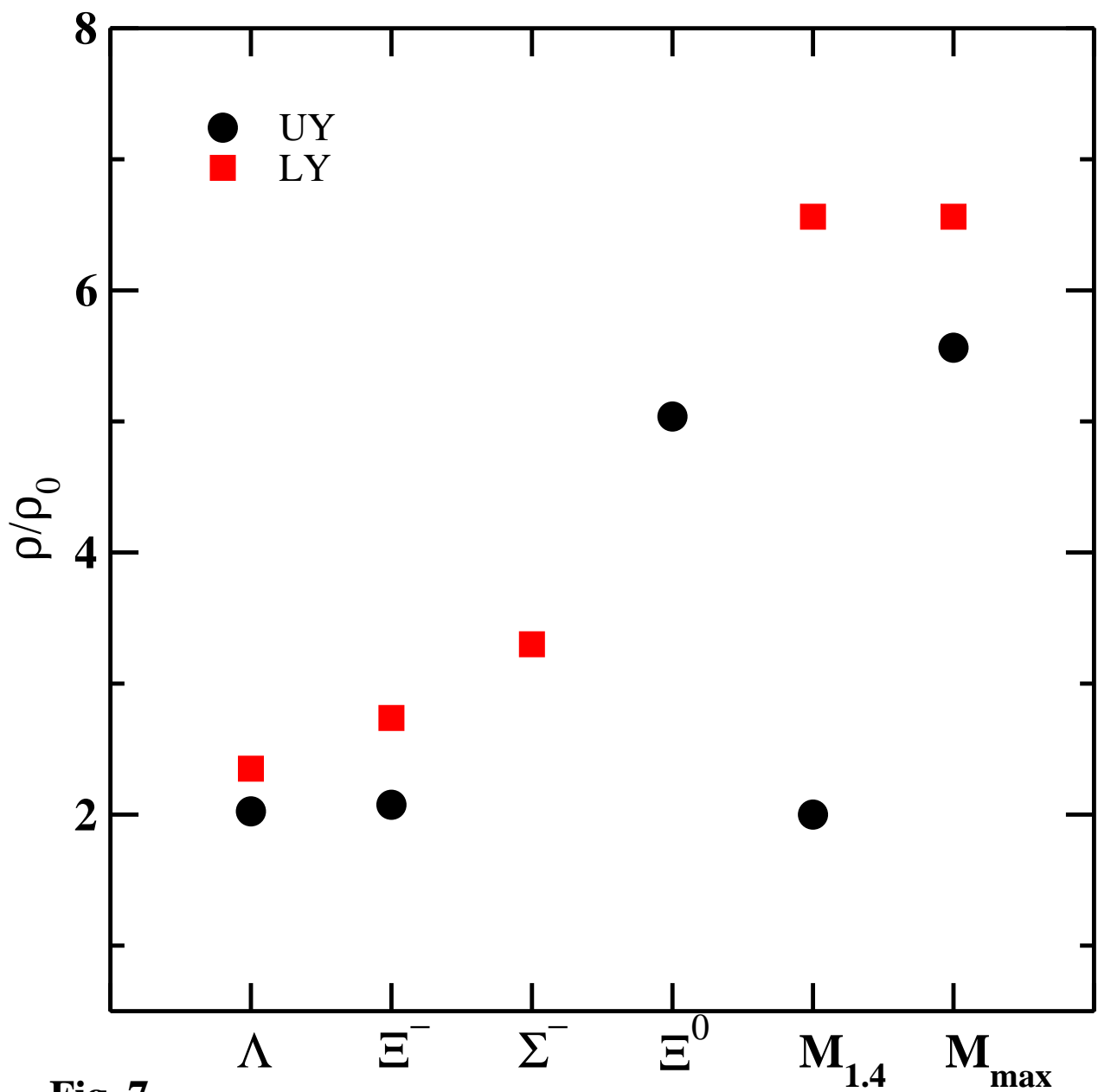
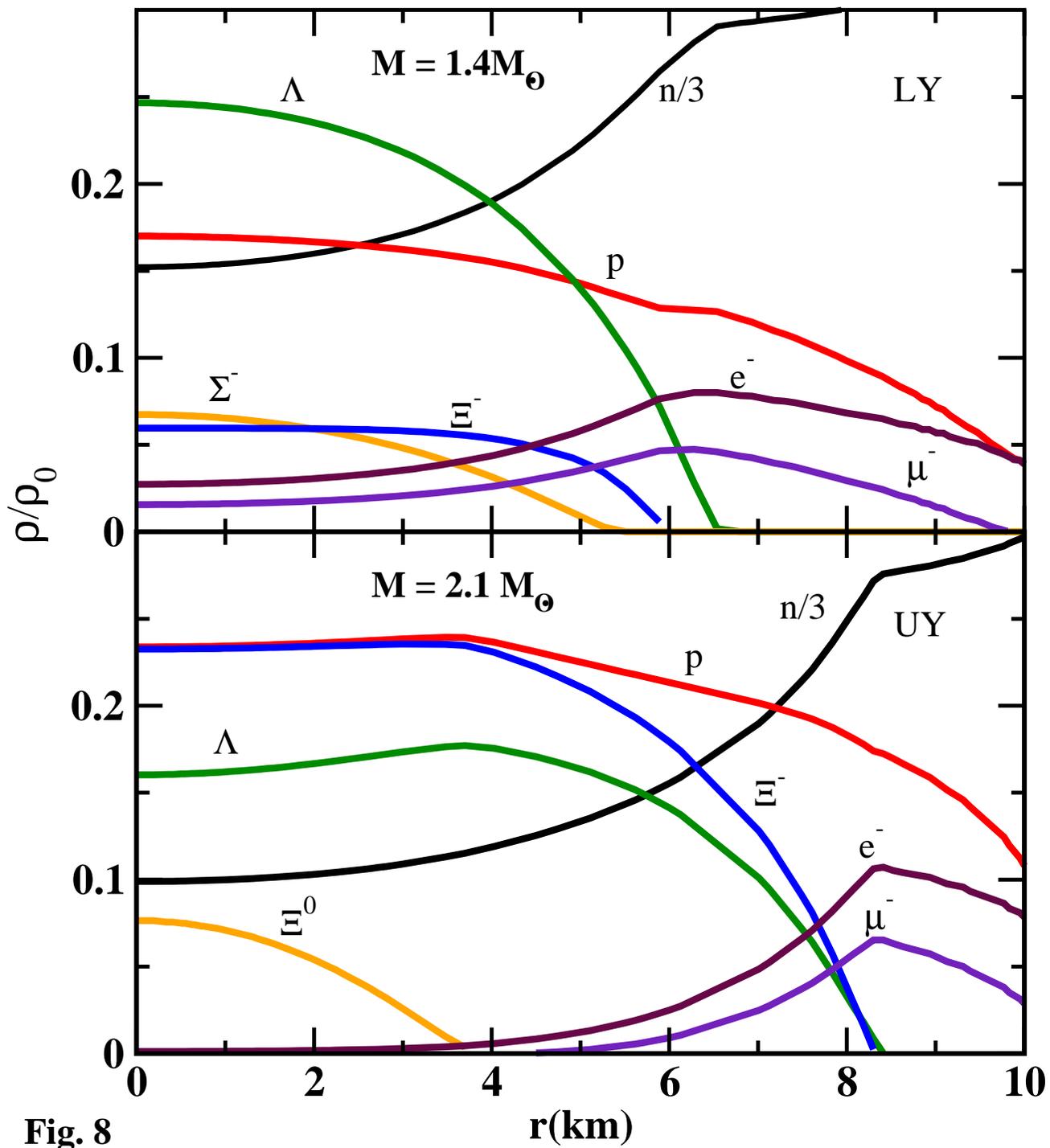
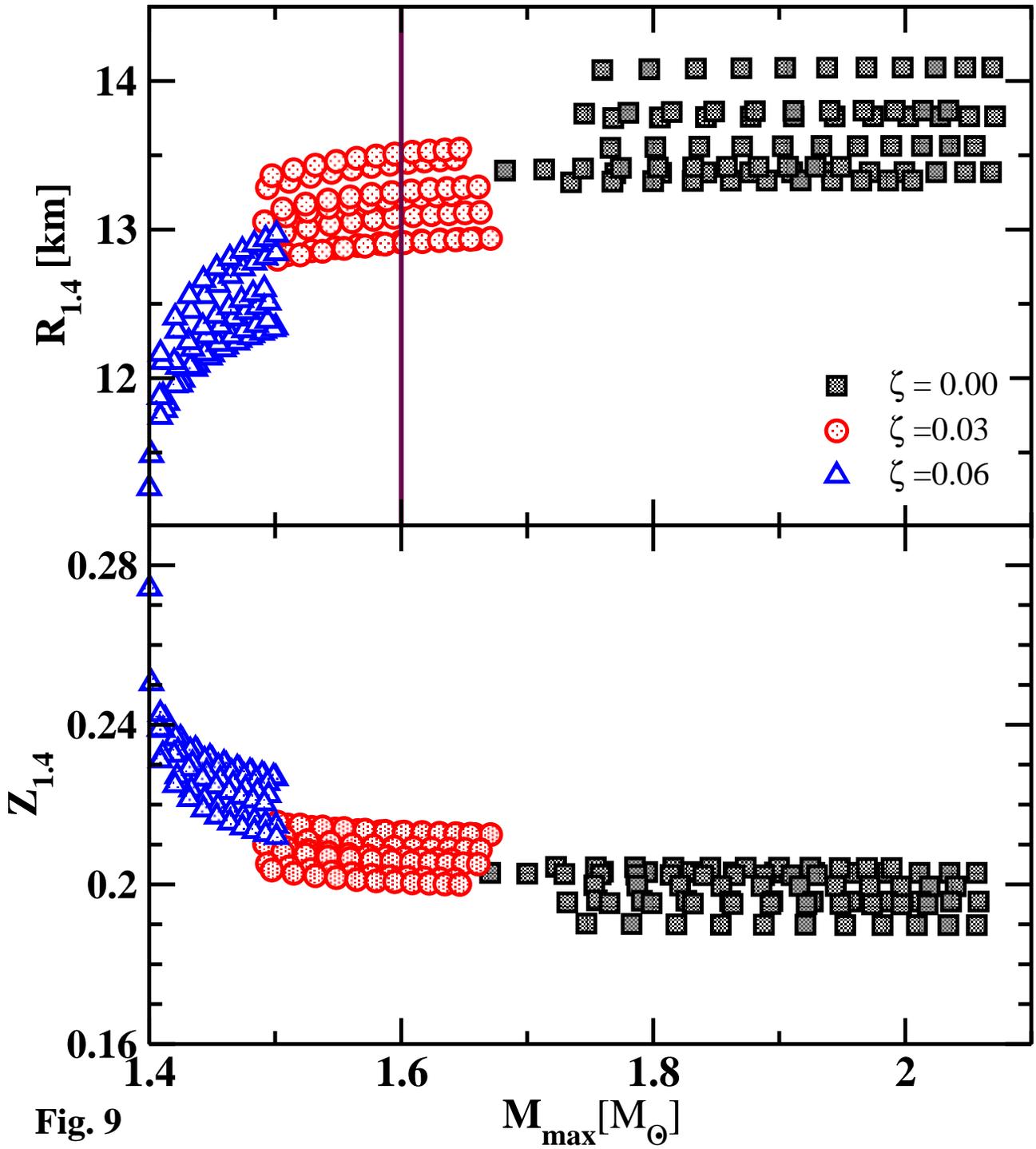
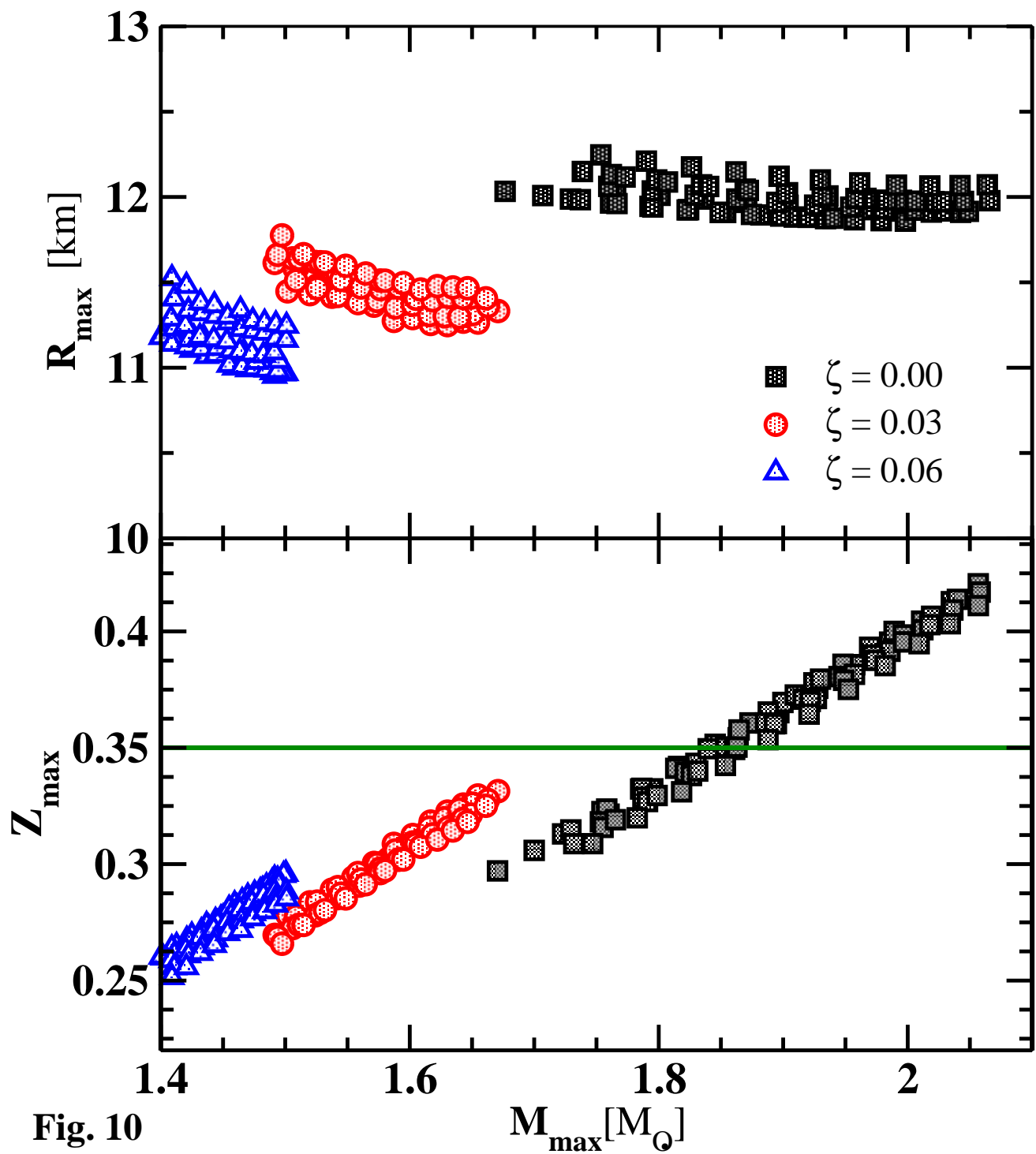
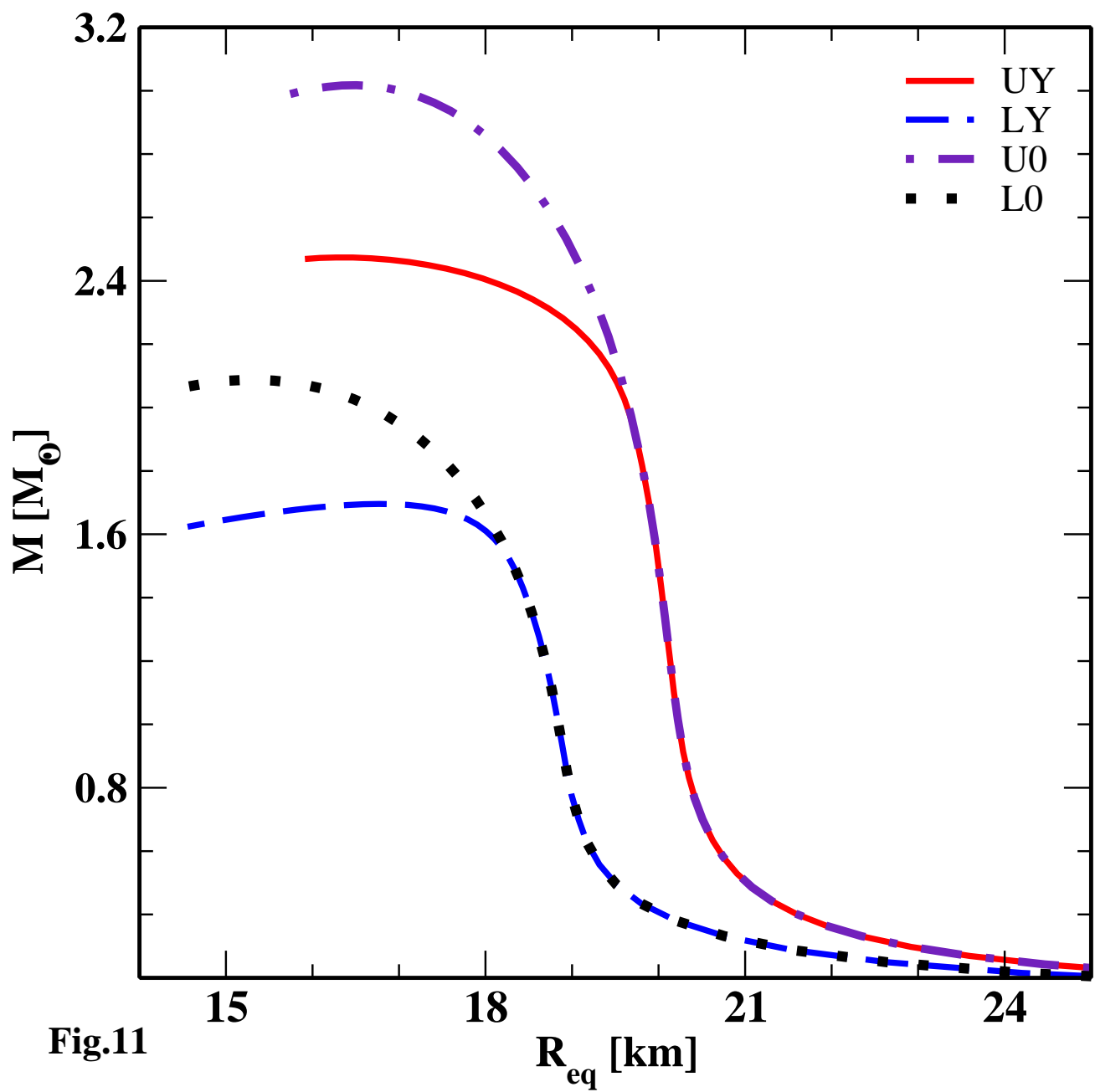
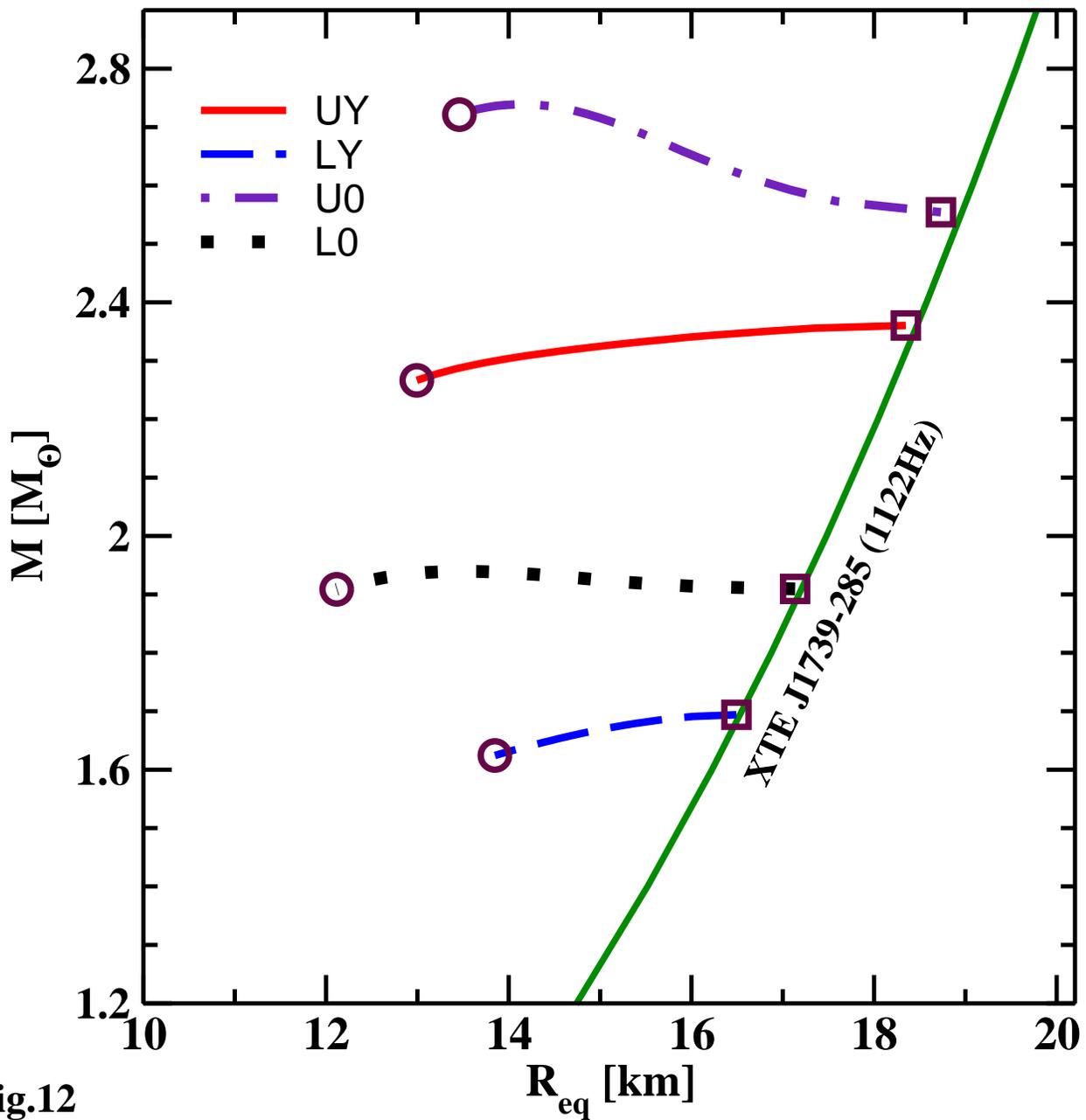
\end{document}